\documentclass[useAMS,usenatbib]{mn2e}
\usepackage{graphicx}
\usepackage{subfigure}
\usepackage{longtable}
\usepackage{fancyhdr}
\def\mag{\hbox{$\stackrel{\raise.3ex\hbox{\scriptsize m}}{\lower.05ex\hbox{.}}$}}               
\def\sech{\hbox{$\stackrel{\raise.3ex\hbox{\scriptsize s}}{\lower.05ex\hbox{.}}$}}          
\def\dgr{\hbox{$\stackrel{\raise.3ex\hbox{\scriptsize $\circ$}}{\lower.05ex\hbox{.}}$}} 
\def\ps{\hbox{s$^{-1}$}}           
\def\sun{_\odot}                                     
\def\rv{radial velocity}              
\def\rvs{radial velocities}              
\def\t{\times}

\title[A study of non-keplerian velocities]{A study of non-keplerian velocities in observations
of spectroscopic binary stars}
\author[J. B. Hearnshaw, S. Komonjinda, J. Skuljan and P. M. Kilmartin]{J. B. Hearnshaw$^{1}$\thanks {E-mail:
john.hearnshaw@canterbury.ac.nz (JBH)}, Siramas
Komonjinda$^{1}$\thanks{E-mail: siramas.k@cmu.ac.th (SK); present
address: Dept. of Physics, Chiang Mai University, Thailand}, J.
Skuljan$^{1}$\thanks{Email: jovan.s@clear.net.nz (JS);
present affiliation: NZ Defence Technology Agency, Auckland}  and P. M. Kilmartin$^{1}$ \\
$^{1}$Dept. of Physics and Astronomy, Univ. of Canterbury,
Christchurch, New Zealand}

\begin{document}

\date{Accepted 2010 April. Received 2010 April; in original form 2010 April}

\pagerange{\pageref{firstpage}--\pageref{lastpage}} \pubyear{2010}

\maketitle

\label{firstpage}

\begin{abstract}
This paper presents an orbital analysis of six southern
single-lined spectroscopic binary systems. The systems selected
were shown to have circular or nearly circular orbits ($e < 0.1$)
from earlier published solutions of only moderate precision. The
purpose was to obtain high precision orbital solutions in order to
investigate the presence of small non-keplerian velocity effects
in the data and hence the reality of the small eccentricities
found for most of the stars.

The Hercules spectrograph and 1-m McLellan telescope at Mt John
Observatory, New Zealand, were used to obtain over 450 CCD spectra
between 2004 October and 2007 August. Radial velocities were
obtained by cross correlation. These data were used to achieve
high precision orbital solutions for all the systems studied,
sometimes with solutions up to about 50 times more precise than
those from the earlier literature. However, the precision of the
solutions is limited in some cases by the rotational velocity or
chromospheric activity of the stars.

The data for the six binaries analysed here are combined with
those for six stars analysed earlier by \citet{sketal:2011}. We
have performed tests using the prescription of \citet{ll:2005} on
all 12 binaries, and conclude that, with one exception, none of
the small eccentricities found by fitting keplerian orbits to the
radial-velocity data can be supported. Instead we conclude that
small non-keplerian effects, which are clearly detectable for six
of our stars, make impossible the precise determination of
spectroscopic binary orbital eccentricities for many late-type
stars to better than about 0.03 in eccentricity, unless the
systematic perturbations are also carefully modelled. The
magnitudes of the non-keplerian velocity variations are given
quantitatively.

\end{abstract}

\begin{keywords}
binaries: spectroscopic; instrumentation: spectrographs;
techniques: radial velocities
\end{keywords}

\section{Introduction}

This paper presents an analysis of six  southern spectroscopic
binary stars which have been analysed using high precision
radial-velocity measurements derived from the cross-correlation of
\'echelle spectra obtained with the Hercules fibre-fed vacuum
spectrograph at Mt John Observatory.

The work continues the programme reported by \citet{sketal:2011}
on single-lined spectroscopic binaries with circular or nearly
circular orbits. The data are used to investigate the possibility
that small non-keplerian velocities have produced a spurious
eccentricity for a binary star with a circularized orbit, as
claimed by \citet{ll:2005}, even for apparent eccentricities as
small as $e \sim 0.01$. The possible nature of non-keplerian
perturbations is discussed in Section \ref{nonkepvel}; in general
they can arise whenever the flux-weighted mean radial velocity of
the observed hemisphere of a star differs from the star's
centre-of-mass radial velocity.

\citet{ll:2005} has shown that such perturbations can in principle
be detected by analysing the amplitude and phase of the third
harmonic in an harmonic analysis of the velocity variations. In
general non-keplerian perturbations, which may affect the second
harmonic, cannot be distinguished from an eccentric keplerian
orbit. However the amplitude and phase of the third harmonic must
be consistent with the second harmonic if the variations are
purely keplerian. Given that the keplerian third harmonic goes as
$e^2$ ($e$ is the eccentricity) it follows that stars with
circular or nearly circular orbits will have very small keplerian
third harmonic terms, which makes it easier to detect a
non-keplerian third harmonic. Our strategy is therefore to select
bright southern late-type SB1 binaries from the literature known
to have small eccentricities from earlier publications.

The idea of non-keplerian perturbations of spectroscopic binary
velocities is far from new, and both \citet{wjl:1936} and
\citet{tes:1941a} considered the likely spurious eccentricities
that would arise. In particular, \citet{tes:1941a} concluded that
the tidal distortion of co-rotating stars could account for this
effect, as it could lead to small line asymmetries and hence
radial-velocity perturbations. \citet{os:1950} also suggested that
spotted stars rotating synchronously with the orbital motion could
result in a similar effect, without giving any quantitative
analysis.

The work of \citet{lbl+mas:1971} devised a statistical test for
the eccentricities of spectroscopic binaries, and concluded that
over 100 of the binary orbits then in the literature with small
non-zero eccentricities in reality failed the test, and their
orbits were really circular. Following the work of
\citet{lbl+mas:1971}, \citet{dgm:1979} emphasized the benefit of
harmonic analysis by determining the second and higher harmonics
in radial velocity curves. He concluded that at least four of the
nine early-type binaries that he analysed have circular orbits, in
spite of a published eccentricity in one case as high as $e=0.09$.

The issue of small eccentricities is re-tackled here, because the
collection of much higher precision data is now possible using
stable spectrographs and CCD detectors, thus making the detection
of apparently significant eccentricities with $e< 0.01$. By
carefully selecting stars known to have nearly circular or
circular orbits, we have been able to show that very small
non-keplerian effects can indeed be detected at the level of $<
10$ m\,\ps\ in the third harmonic.

The present programme selected six spectroscopic binaries from the
{\sl Ninth catalogue of spectroscopic binary orbits}, $S_{B^9}$
\citep{dpetal:2004}, which contains over 3300 orbital solutions
for 2770 binaries. The selection criteria were that the earlier
analysis reported an eccentricity $e < 0.1$, with F, G or K
spectral type for the primary star, with a southern declination
($\delta < 0^{\circ}$) and brighter than $m_V = 8.0$. In addition
systems with rather long periods ($P > 850$~d) or very short
periods ($P < 2.2$ d) were also excluded, and three stars with
high levels of chromospheric activity and large $v\sin i$ were
also not included. Only 11 binary systems satisfied all these
selection criteria. One additional system, HD\,101379 (GT Muscae),
which was strangely not listed in $S_{B^9}$, but which otherwise
satisfies the criteria, was also included in this programme. Its
orbit was analysed earlier by \citet{kametal:1995}. This paper
reports the results for six systems (including GT Muscae).
Together with those analysed by \citet{sketal:2011} this brings
the total to 12 systems selected, being all those satisfying the
criteria. All are late-type evolved stars.

Table 1 lists the six objects selected for observation and
analysis in this paper.

\begin{table*}
\centering
    \caption[Selected binary systems studied in this paper]{The six selected spectroscopic binaries studied
    in this paper. All the stars are listed
    in $S_{B^9}$ except for HD\,101379.}
\begin{tabular}{ r c c c l l }
\hline

\textbf{HD}  &  \textbf{RA(2000.0)}  &  \textbf{Dec(2000.0)}  &
\textbf{$m_V$} &  \textbf{Spectral type}  &  \textbf{$P$} (days) \\
\hline
 77258  & 09 00 05.44 & --41 15 13.5 & 4.45 & G8-K1III    &  74.1469  \\
 85622  & 09 51 40.69 & --46 32 51.5 & 4.57 & G5Ib                    & 329.3     \\
101379  & 11 39 29.59 & --65 23 51.9 & 5.17 & G2III & 61.448       \\
124425  & 14 13 40.67 & --00 50 42.4 & 5.93 & F6IV                &   2.696   \\
136905  & 15 23 26.06 & --06 36 36.7 & 7.31 & K1III           &  11.1345  \\
194215  & 20 25 26.82 & --28 39 47.8 & 5.84 & G8II/III      & 377.6         \\
    \hline
        \end{tabular}
  \label{tab:list}
\end{table*}

\section{Observations and statistics}

The observational part of this research was carried out at Mt John
University Observatory (New Zealand) from 2004 October to 2007
August. All observations were carried out on the 1-m McLellan
telescope, using the Hercules fibre-fed vacuum \'echelle
spectrograph  \citep{jbhetal:2002}. Relevant details of the
observing procedure are given by \citet{sketal:2011}. For all
stars a resolving power of 70\,000 was used, except for
HD\,136905, where $R=41\,000$ was adopted, as this star is
somewhat fainter than the others. A typical signal-to-noise ratio
of $\sim 100:1$ was obtained for our spectra in a pixel-column.

The Hercules detector mainly used was a SITe SI\,003
1024$\times$1024 thinned CCD chip with 24-micron square pixels.
(In \citet{sketal:2011} this was incorrectly given as 23-micron
pixels.) This chip cannot cover all the focal plane area of the
dispersed spectrum, but was positioned so as to cover 46 orders
from 457 to 722 nm (orders 79 to 124). Later in 2006, the detector
for Hercules was changed to a Spectral Instruments camera with a
Fairchild 486 CCD which has 4096$\times$4096 15-micron square
pixels. This was only used for two runs in 2007 August.

\section{Data reduction and radial-velocity determination}

All the spectra obtained were reduced using the Hercules Reduction
Software Package, HRSP version 2.3 \citep{js:2003}. This software
is a C program running under Linux. The dispersion solution was
obtained from two thorium-argon lamp spectra, recorded immediately
before and after each stellar spectrum. About 400 Th or Ar lines
were used. Further details are given by \citet{sketal:2011}.

All radial velocities were determined by cross correlation for the
30 \'echelle orders used, using one chosen spectrum of the same
star as a template relative to which all other spectra of that
star were measured, and following the Fourier cross-correlation
techniques pioneered by \citet{sms:1974} for digital spectra. In
particular, the cross-correlation was performed in $\ln\lambda$
space after subtracting the mean flux from each spectrum and
applying a cosine bell to the ends of the data window. The
position of the cross-correlation function (CCF) peak was
determined by means of a gaussian least squares fit, typically
using eight consecutive data points spanning the peak.

\section{Orbit analysis from radial velocities}

The six orbital elements of a particular binary system, $K$
(radial-velocity semi-amplitude), $e$ (orbital eccentricity),
$\omega$ (longitude of periastron), $T_0$ (time of zero mean
longitude), $P$ (period) and $\gamma$ (the centre-of-mass radial
velocity), can be determined from a set of $n$ observations of
radial velocity $V_{\mathrm{rad}}(t)$. The best values of these
elements can be found following an iterative least-squares
analysis involving differential corrections, based on the method
of \citet{rlf:1894}. Data points lying more than $3 \sigma$ from
the fitted solution have been rejected and the remaining points
then used for a second solution; rejected points are still plotted
in the figures and they appear in the tables of measured radial
velocities.

Here we use a software package SpecBin written by J.\ Skuljan
\citep{jsetal:2004}. Since the orbits here are close to being
circular, the variation of the Lehmann-Filh\'es procedure
described by \citet{tes:1941b} was adopted, unless otherwise
stated. This makes use of the time of zero mean longitude ($T_0$)
instead of the poorly determined time of periastron passage ($T$).
The mean longitude is given by $L=2\pi(t-T)/P + \omega$ and hence
the time when $L$ is zero is $T_0 = T -\omega P/2\pi$. In fact,
for one star there is a well-determined non-zero eccentricity
(HD\,194215), so for this star we were able reliably to determine
the time of periastron passage, and hence we used that time for
the phase zero-point.

Since our \rvs\ are relative to a template observation of the same
star, our orbital solutions quote $\gamma_{\rm rel}$. In addition
our solutions from Hercules data give an absolute centre-of-mass
velocity, $\gamma$, of lower accuracy, based on cross-correlating
the template with a standard star. Further details of SpecBin and
the process of estimating the error bars in the orbital elements
are discussed by \citet{jsetal:2004}.

We have also used SpecBin to reanalyse the historical data for the
stars in the literature. This has allowed us to compare $T_0$
values from the historical data with those from our data, and
hence to derive better periods (for four of the stars) than from
either data set alone, using the long time base line.

\section{Results of the analysis for individual stars}

In this section the results of the orbital analysis of each of the
binary systems are presented.

\subsection{HD\,77258}

The bright star HD\,77258 (w Velorum, HR\,3591) was first observed
to have a variation in radial velocity by H.K.\ Palmer, as
reported in the Lick Observatory Bulletin in 1904
\citep{whw:1904}.


The spectral type was classified as F8IV by \citet{adv:1957}, but
it is given as G8-K1III+A in the Michigan Spectral Catalogue
\citep{nh:1982}. The Hipparcos photometry of HD\,77258 shows that
no detectable light variation was present. The parallax of this
system in the Hipparcos main catalogue is $16.19 \pm 0.53$ mas,
which implies a distance of $61.8 \pm 2.0$ pc. The absolute
magnitude is $M_V = +0.50$, supporting the giant luminosity class.

\citet{jl:1919} observed five spectra of HD\,77258 at the Cape
during 1914--1916. He showed that the star has a range in radial
velocity of 34.2 km\,\ps. During February and May 1921, 38 further
spectra were observed and the radial velocities were used for an
orbital solution of low precision (root mean square residual of
1.23 km\,\ps), as reported by \citet{jl:1924}.

The Hercules radial velocities for HD\,77258 were measured from
101 spectra and are given in Table \ref{rv:HD77258}. The orbital
solution found has an rms scatter of 19 m\,\ps, which is
comparable with the lowest scatter in \rvs\ for sharp-lined stars
observed with Hercules.

\begin{figure}
    \centering
        \subfigure[Phase plot of HD\,77258 \rvs.]{\includegraphics[width=9.25cm]{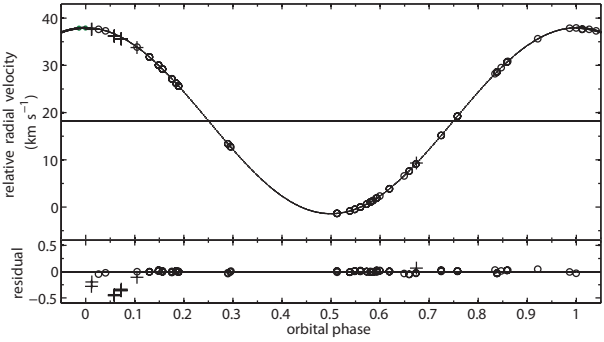}}\\
               \subfigure[Measured \rvs\ of HD\,77258 versus Julian date.]{\includegraphics[width=9.25cm]{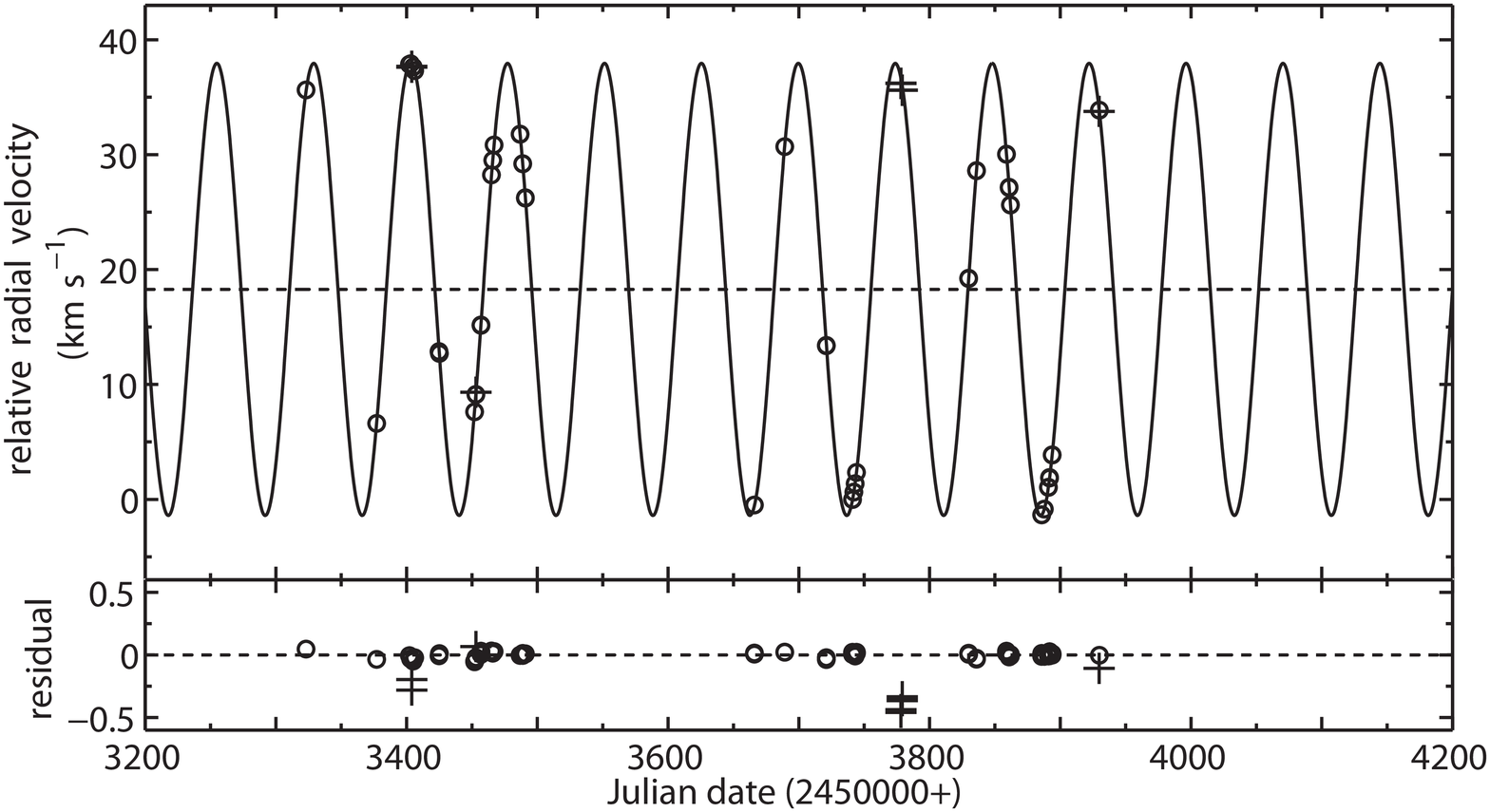}}\\
    \caption[Radial velocity curves of HD\,77258]{Radial velocities of HD\,77258 measured from the
    Hercules spectra. The plotted curves are calculated from the orbital solution calculated from these \rvs\
    with all free parameters. The second panel of each figure shows the residuals from the fit. Open circles are points
    used in the orbital solution, and the crosses (+) are rejected points with residuals of at least $3\sigma$.}
    \label{fig:HD77258}
\end{figure}

The orbital parameters are presented in Table \ref{tab:HD77258}.
Figure \ref{fig:HD77258} is a plot of these radial velocities as a
function of (a) orbital phase and (b) Julian Date. The orbital
phase is calculated from the period of 74.13715 days and the zero
phase is defined by $T_0 = \mathrm{JD}\,245\,3625.5112 \pm
0.0017$, the time of zero mean longitude.
\begin{table}
    \caption[Orbital solutions of HD\,77258 from Hercules data.]{The orbital parameters
    of SB1 HD\,77258 (F8IV) as obtained from Hercules data.}
    \label{tab:HD77258}
    \begin{center}
        \begin{tabular}{ l l l}
            \hline
   \textbf{Parameter}   & \multicolumn{2}{c}{\textbf{Hercules solutions}} \\
                & \small{no fixed parameter}    & \small{when $P$ = $\bar{P}$} \\
            \hline
   \textit{K} (km\,\ps)         & 19.6744 $\pm$ 0.0041  & 19.6747 $\pm$ 0.0039  \\
   \textit{e}           & 0.000\,85 $\pm$ 0.000\,19 & 0.000\,87 $\pm$ 0.000\,19 \\
   \textit{$\omega$} ($^\circ$) & 106 $\pm$ 13  & 106 $\pm$ 12              \\
   \textit{$T_0$} (HJD) & 245\,3625.5112$\pm$ 0.0017 & 245\,3625.5113 $\pm$ 0.0017  \\
   \textit{P} (days)        & 74.137\,15 $\pm$ 0.000\,73 & 74.1368 (fixed)                  \\
   \textit{$\gamma_{\mathrm{rel}}$} (km\,\ps)   & 18.2758 $\pm$ 0.0027 & 18.2757 $\pm$ 0.0027       \\
   \textit{$\gamma$} (km\,\ps)  & $-$5.0133 $\pm$ 0.042 & $-$5.0134 $\pm$   0.042                           \\
   \textit{$\#_{obs}$}                  & 101            & 101                                                    \\
   \textit{$\#_{rej}$}                  & 10             & 10                                                     \\
   \textit{$\sigma$} (km\,\ps)  & 0.019     & 0.019                                             \\
   \textit{$a_1\sin i$} ($\t 10^7$ km)  & $2.0057 \pm 0.0004$ & $2.0057 \pm 0.0005$      \\
   \textit{$f(M)$} ($M_{\odot}$) & $0.058498 \pm 0.000037$ & $0.058500 \pm 0.000037$ \\
        \hline
        \end{tabular}
    \end{center}
\end{table}

The period calculated from $T_0$ from the combined data of
\citet{jl:1924} and from the Hercules data is $\bar{P} = 74.1368
\pm 0.0023$ days. This period has a lower precision than the
period that was calculated from the Hercules data alone (in spite
of about eight decades of baseline between the two times of data
collection). Table \ref{tab:HD77258} also shows the solution with
the period fixed using the combined data set; it differs
negligibly from the solution from Hercules data alone.

\subsection{HD\,85622}

HD\,85622 (m Velorum, HR\,3912) is a system with a supergiant
primary star. It was first found to have variable radial velocity
in 1905 during the D.O.\ Mills Expedition \citep{whw:1907}. It was
classified as a supergiant star of spectral type G5Ib by
\citet{djm+wpb:1976}. The rotational velocity, $v\sin i$, of the
supergiant component was measured by \citet{jrdmetal:2002} as
$19.3 \pm 1.0$ km\,\ps.


The Hipparcos and Tycho photometry indicated that this system has
a micro-variability in its brightness with 0\mag007 scatter in
$H_p = 4\mag749 \pm 0\mag0007$ and 0\mag023 scatter in $V_T =
4\mag713 \pm 0\mag002$. In other words the intrinsic variability
is about ten times larger than the standard error of photometric
measurement.


\citet{wjl:1936} collected 56 radial-velocity observations of
HD\,85622 from the papers by \citet{jl:1919} and \citet{hsj:1928}.
He adopted a circular orbit for this system with an rms residual
of 1.4 km\,\ps.

Eighty-six radial velocities were measured from the Hercules
spectra obtained. It was found that the calculated eccentricity
has a large error bar, $e = 0.0026 \pm 0.0016$. A circular orbit
should therefore be adopted from these data with a precision of 61
m\,\ps. These two solutions, with the eccentricity floating or
fixed to zero, can be compared in Table \ref{tab:HD85622}.

\begin{figure}
 \centering
\subfigure[Phase plot of HD\,85622 \rvs.]{\includegraphics[width=9.25cm]{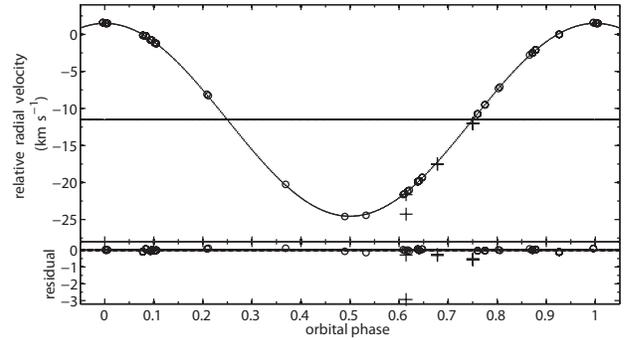}}\\
\subfigure[Measured \rvs\ of HD\,85622 versus Julian
date.]{\includegraphics[width=9.25cm]{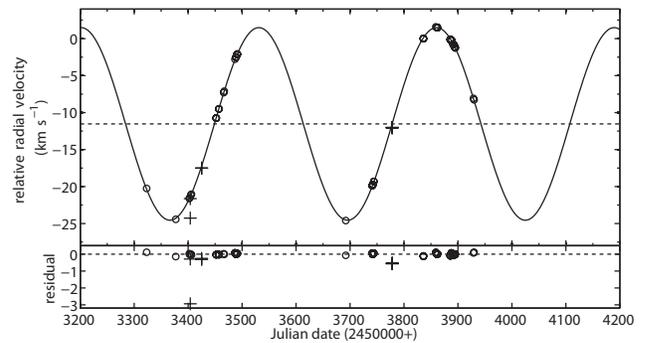}}

    \caption[Radial velocity curves of HD\,85622]{Radial velocities of HD\,85622 measured from the Hercules
    spectra. The plotted curves are calculated from the circular orbital
solution. The second panel shown the residuals from this fit.}
    \label{fig:HD85622}
\end{figure}

The radial velocities of this system and the residuals for the
circular orbital solution are plotted versus orbital phase in
Fig.\ \ref{fig:HD85622} (a) and versus Julian date in Fig.\
\ref{fig:HD85622}(b). The zero phase is at the time $T_0 =
\mathrm{JD}\,245\,3860.281 \pm 0.074$.

\begin{table}
    \caption[Orbital solutions of HD\,85622 from Hercules data.]{The orbital parameters of
    SB1 HD\,85622 analysed from Hercules data.}
    \label{tab:HD85622}
    \begin{center}
        \begin{tabular}{ l l l }
            \hline
   \textbf{Parameter}   &  \multicolumn{2}{c}{\textbf{Hercules solutions}} \\
                                        & \small{eccentric orbit} & \small{circular orbit}\\
            \hline
   \textit{K} (km\,\ps) & 13.028 $\pm$ 0.013            & 13.021 $\pm$ 0.012    \\
   \textit{e}                   & 0.0026 $\pm$ 0.0016       & 0 (adopted)                   \\
   \textit{$\omega$} ($^\circ$) & 188 $\pm$ 46      & --                                     \\
   \textit{$T_0$} (HJD) & 245\,3860.261 $\pm$ 0.078 & 245\,3860.281 $\pm$ 0.074 \\
   \textit{P} (days)        & 329.126 $\pm$ 0.093       & 329.266 $\pm$ 0.085 \\
   \textit{$\gamma_{\mathrm{rel}}$} (km\,\ps)   & $-$11.516 $\pm$ 0.013 & $-$11.531 $\pm$ 0.010 \\
   \textit{$\gamma$} (km\,\ps)  & 11.510 $\pm$ 0.037 & 11.495 $\pm$ 0.034 \\
   \textit{$\#_{obs}$}  & 86                                            & 86                                    \\
   \textit{$\#_{rej}$}  &  6                                            & 6                                     \\
   \textit{$\sigma$} (km\,\ps)  & 0.060                     & 0.061                             \\
   \textit{$a_1\sin i$} ($\t 10^7$ km)  & $5.8962 \pm 0.0076$ & $5.8956 \pm 0.0076$      \\
   \textit{$f(M)$} ($M_{\odot}$) & $0.07540 \pm 0.00025$ & $0.07532 \pm 0.00023$ \\
        \hline
        \end{tabular}
    \end{center}
\end{table}

\subsection{HD\,101379}

HD\,101379 or GT Muscae is a member of a quadruple system. The
brightness is variable, because the system comprises an RS
CVn-type single-lined spectroscopic binary, HD\,101379 (orbital
period $\sim 61.4$ d) and the eclipsing binary, HD\,101380
(eclipsing period $\sim 2.75459$ d). \citet{hmetal:1990}
reported the separation between the two components as 0.23 arcsec
from speckle interferometry.

The spectral types of the stars in this system have been
classified by many authors. \citet{nh+apc:1975} 
have classified HD\,101379 as G5/G8III and HD\,101380 as A0/1V.
\citet{acc:1982} 
determined the spectral types of HD\,101379 as K4III and of the
components of the eclipsing system as A0V and A2V from his
photometric analysis. 

The RS CVn-type system HD\,101379 shows a strong CaII H and K
emission and a variable H$\alpha$ line \citep{accetal:1982}.




An analysis of GT\,Mus spectra was done by \citet{kametal:1995}.
In that paper, the orbital solution of the system HD\,101379 was
analysed from the 17 spectra that were obtained from the MJUO's
Cassegrain \'echelle spectrograph linked with the McLellan 1-m
telescope, which were combined with other radial velocities
reported by \citet{lab:1987}, 
\citet{acc:1987} 
and \citet{pjm:1986}. 

In this study, 76 high resolution spectra of GT Mus at a
wavelength 450--720 nm were obtained from the Hercules
spectrograph. These spectra, which are dominated by HD\,101379,
were obtained over a period of two years (2004 October -- 2006
June) with a signal-to-noise ratio of typically 70. Relative
radial velocities of these spectra were calculated 
using the cross-correlation technique with a template spectrum.
These relative \rvs\ of HD\,101379 can be converted into absolute
\rvs\ using the relative radial velocity of the template spectrum
with respect to a standard star spectrum. The spectrum of the
standard star HD\,109379 (G5II) was used for this purpose. The
relative radial velocity of the template with respect to this
standard star is $21.110 \pm 0.024$ km\,\ps. The standard star
absolute velocity is $-7.246 \pm 0.024$ km\,\ps, as obtained by
\citet{djr:2004}.

The orbital solution was calculated using the above radial
velocities. The solution, as in Table \ref{tab:HD101379}, has an
rms of 518 m\,\ps. This is three times larger than the precision
of the \citet{kametal:1995} radial velocities, for the reasons
discussed below.

\begin{table}
    \caption{The orbital parameters of SB1 HD\,101379 (G5/8III) derived from Hercules \rvs, and
    from Hercules data combined with previously published radial velocities.}
    \label{tab:HD101379}
    \begin{center}
        \begin{tabular}{ l l l }
            \hline
   \textbf{Parameter} &  \textbf{New analysed values} & \textbf{Combined data}    \\
                      &  Hercules data only          &                           \\
   \hline
   \textit{K} (km\,\ps)                 & 12.911 $\pm$ 0.078                & $12.823 \pm 0.049$    \\
   \textit{e}                           & 0.012 $\pm$ 0.010                 & $0.0053 \pm 0.0034$    \\
   \textit{$\omega$} ($^\circ$)         & $237 \pm 51$                      & $73 \pm 51$            \\
   \textit{$T_0$} (HJD)                 &  245\,3736.00 $\pm$ 0.10          & $244\,4951.03 \pm 0.13$  \\
   \textit{P} (days)                    & 61.408 $\pm$ 0.027                & $61.4370 \pm 0.0012$     \\
   \textit{$\gamma$} (km\,\ps)          & 1.02 $\pm$ 0.11                   &     \\
   \textit{$\#_{obs}$}                  & 76                                & 183            \\
   \textit{$\#_{rej}$}                  & 0                                 & 1   \\
   \textit{$\sigma$} (km\,\ps)          & 0.518                             & 0.380         \\
   \textit{$a_1\sin i$} ($\t 10^7$ km)  & $1.0902 \pm 0.0071$               & $1.0833 \pm 0.0042$   \\
   \textit{$f(M)$} ($M_{\odot}$)        & $0.01369 \pm 0.00025$             & $0.01342 \pm 0.00015$    \\
        \hline
        \end{tabular}
   \end{center}
\end{table}

\begin{figure}
    \centering
        \subfigure[Phase plot of HD\,101379 \rvs.]{\includegraphics[width=9.25cm]{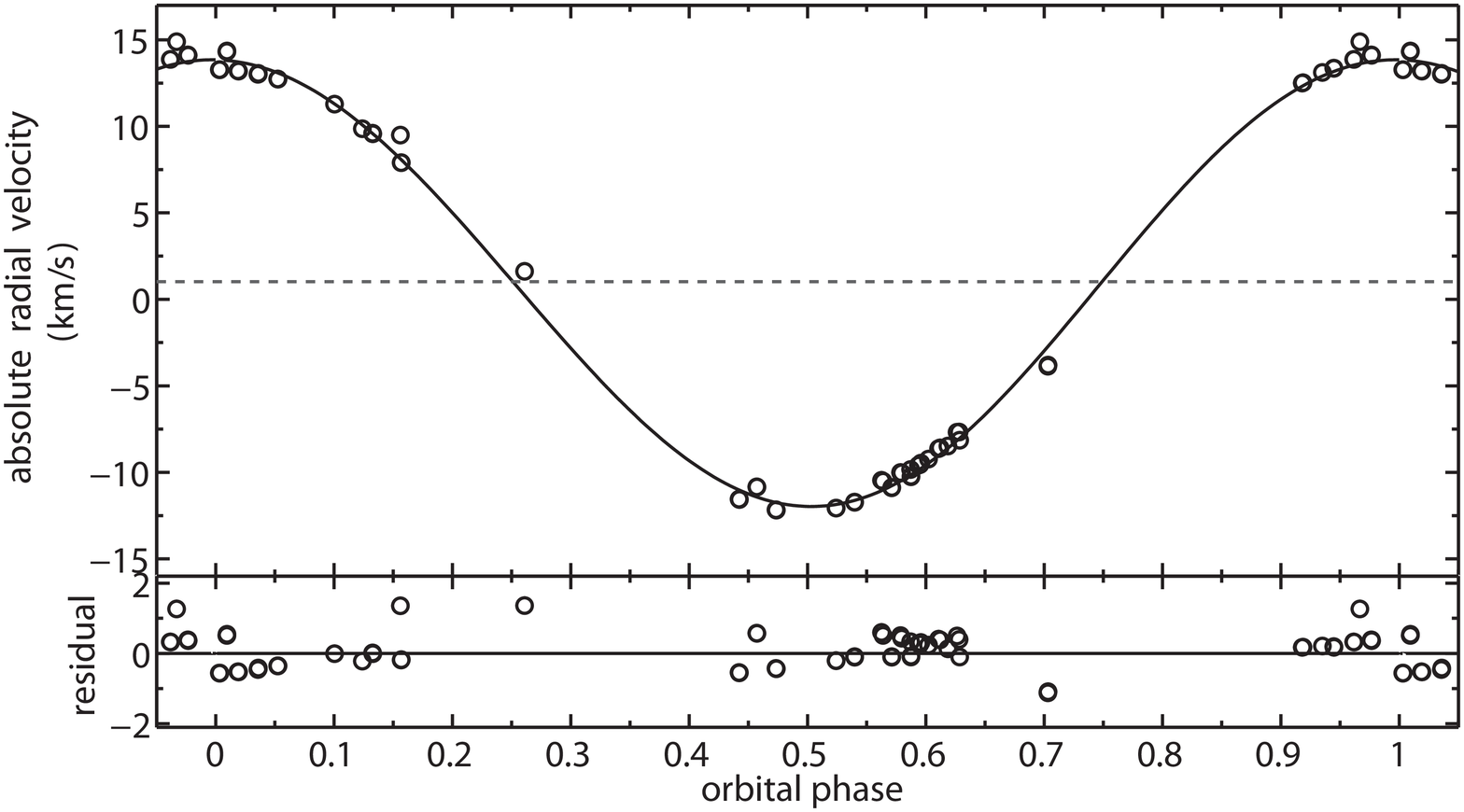}}\\
        \subfigure[Measured \rvs\ of HD\,101379 versus Julian date.]{\includegraphics[width=9.25cm]{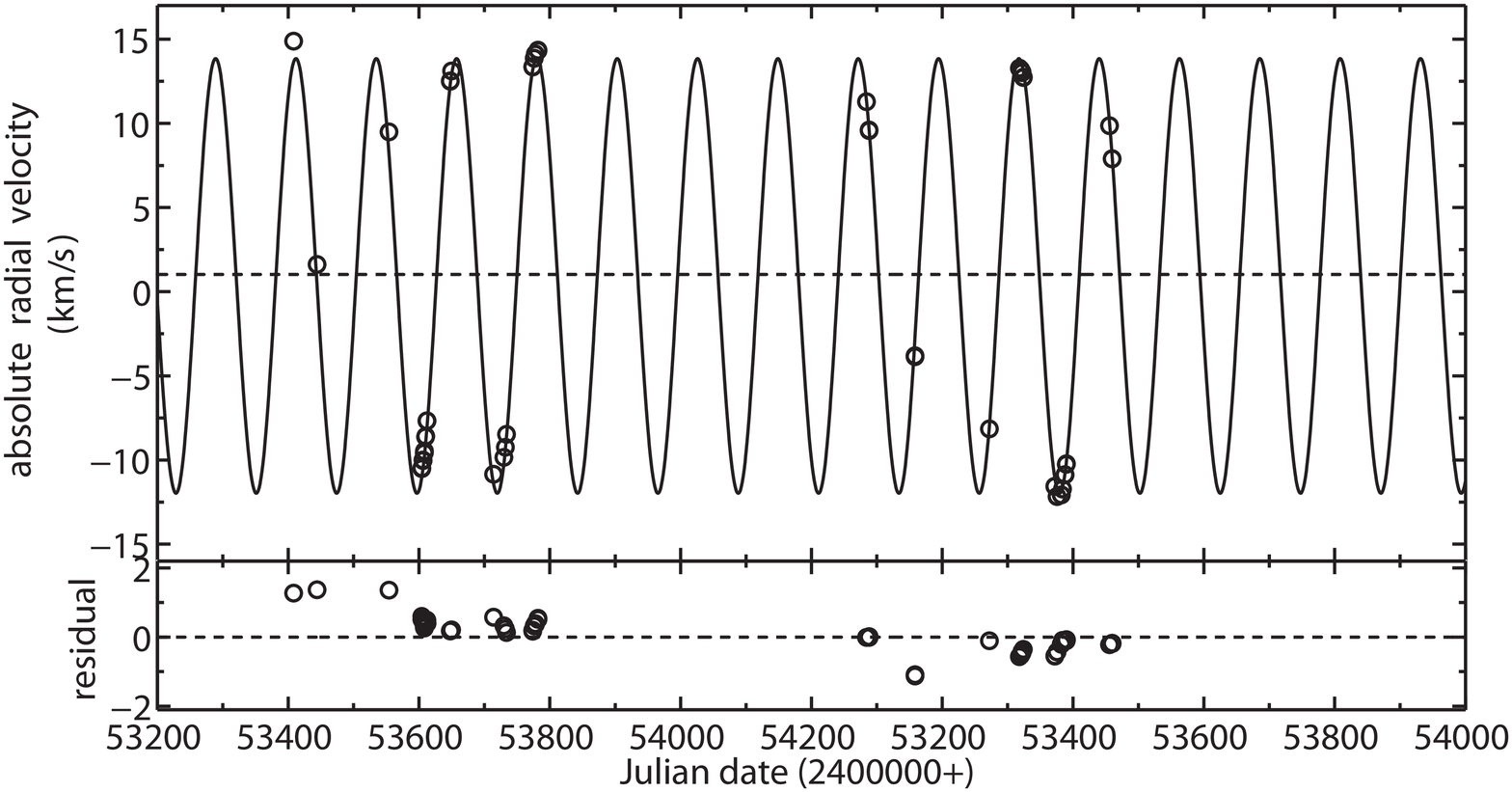}}
    \caption[Radial velocity curves of HD\,101379]{Radial-velocity observations of GT Mus measured from Hercules spectra.
    The plotted curves are calculated from the orbital solution calculated from these \rvs. The second panels
    show the residuals from the fits.}
    \label{fig:HD101379_new}
\end{figure}

Fig.\ \ref{fig:HD101379_new}a and b show the relative radial
velocities measured from the Hercules spectra with their residuals
in the second panel. It is clearly seen that the residuals from
this fit have a positive value during the first observation period
and have a negative value during the second period. This should be
due to the fact that the system GT\,Mus is a quadruple system,
with HD\,101379 showing long-period orbital motion with its
companion, HD\,101380. This long scale variation was not shown in
the \rvs\ of \citet{kametal:1995} because of their shorter
observation window (JD 244\,8260 -- 244\,8479).

When these data are separated into two groups by the observation
time (2004 October -- 2005 May and 2005 November -- 2006 June),
each data set gave a better solution, with rms residuals of 79
m\,\ps\ and 54 m\,\ps\ respectively, as shown in Table
\ref{tab:HD101379_part}.
\begin{table*}
    \caption[HD\,101379 orbital solutions from two period of observations]{The orbital parameters of HD\,101379
    derived from two data sets of Hercules \rvs, 2004 October -- 2005 May and 2005 November -- 2006 June.}
    \label{tab:HD101379_part}
    \begin{center}
        \begin{tabular}{l l l}
            \hline
   \textbf{Parameter} & \multicolumn{2}{c}{\textbf{Hercules solutions}} \\
                                        & \small{2004 October -- 2005 May} & \small{2005 November -- 2006 June} \\
   \hline
   \textit{K} (km\,\ps) & 12.948 $\pm$ 0.027                & 12.715 $\pm$ 0.011 \\
   \textit{e}                   & 0.0225 $\pm$ 0.0038           & 0.0331 $\pm$ 0.0027 \\
   \textit{$\omega$} ($^\circ$) & 175.3 $\pm$ 6.9   & 61.8 $\pm$ 2.1\\
   \textit{$T_0$} (HJD) & 245\,3429.478 $\pm$ 0.036 & 245\,3797.763 $\pm$ 0.018 \\
   \textit{P} (days)        & 61.516 $\pm$ 0.045                & 61.331 $\pm$ 0.017 \\
   \textit{$\gamma$} (km\,\ps)  & 1.723 $\pm$ 0.035 & 0.287 $\pm$ 0.026 \\
   \textit{$\#_{obs}$}                  & 34                & 42            \\
   \textit{$\#_{rej}$}                  & 1                 & 1             \\
   \textit{$\sigma$} (km\,\ps)  & 0.079         & 0.054     \\
   \textit{$a_1\sin i$} ($\t 10^7$ km)  & $1.0950 \pm 0.0031$ & $1.0717 \pm 0.0017$   \\
   \textit{$f(M)$} ($M_{\odot}$) & $0.01383 \pm 0.00010$  & $0.01304 \pm 0.00006$ \\
            \hline
        \end{tabular}
    \end{center}
\end{table*}

The orbital solutions from both data sets do not change
dramatically, except the longitude of periastron $\omega$ and the
systemic radial velocity $\gamma$. The value of $\omega$ changed
from $175\dgr3 \pm 6\dgr9$ to $61\dgr8 \pm 2\dgr1$ and the value
of $\gamma$ changed from $1.723 \pm 0.035$ to $0.287 \pm 0.26$
km\,\ps. These changes result respectively from the apsidal motion
of HD\,101379 caused by its companion system, HD\,101380, and by
the orbital motion of the visual pair around the centre of mass of
the GT Mus quadruple system.


To improve the solution, the historical \rvs\ were analysed
together with Hercules \rvs. The \rvs\ of HD\,101379 were observed
and reported by several observers using different instruments.
These include \citet{pjm:1986}, 
\citet{acc:1987} 
and \citet{kametal:1995}. 
The other observations of HD\,101379 that are included in this
analysis are from \citet{lab:1987}. 
The data from MacQueen (1986) could not be combined into this
analysis, as his radial velocities showed an orbital period of
around 54 days. The orbital solutions were calculated from all the
other data sets.

\begin{table}
    \caption[Systemic radial velocity of HD\,101379]{The systemic radial velocity $\gamma$ of HD\,101379
    calculated from historical \rvs\ and Hercules \rvs.}
    \label{tab:HD101379_gamma}
    \begin{center}
        \begin{tabular}{l c c}
            \hline
   \textbf{Radial velocity data}    & \textit{$T_0$}            & \textit{$\gamma$} \\
                                                        &   \small{(HJD 2400000+)}  & \small{(km\,\ps)} \\
            \hline
   Balona (1987)                    & 44\,397.25 $\pm$ 0.63     & 9.12 $\pm$ 0.50 \\
   Collier Cameron (1987)   & 44\,458.71 $\pm$ 0.59     & 8.13 $\pm$ 0.38 \\
   Murdoch et al.\ (1995)                   & 48\,391.76 $\pm$ 0.42     & 0.53 $\pm$ 0.47 \\
   Hercules data set 1      & 53\,429.478 $\pm$ 0.036 & 1.723 $\pm$ 0.035 \\
   Hercules data set 2      & 53\,797.763 $\pm$ 0.018 & 0.287 $\pm$ 0.026 \\
            \hline
        \end{tabular}
    \end{center}
\end{table}

The systemic radial velocities, $\gamma$, and times of zero mean
longitude, $T_0$, of each data set are shown here in Table
\ref{tab:HD101379_gamma}. These $\gamma$ values decrease during 30
years of observation. The orbital period of HD\,101379 around the
centre-of-mass of GT Mus cannot be calculated from these few data
points. From the Hipparcos catalogue, the parallax is $5.81 \pm
0.64$ mas and the separation between the two stars is $0.217 \pm
0.004$ arcsec. The latter value is consistent with the
interferometric measurement of \citet{hmetal:1990}.
The HIP astrometry also measured the changing rate of the position
angle between the two systems as $d\theta/dt = 3\dgr0$ per year.
This information gave a lower limit of the total system mass as
$3.6M_{\sun}$. This value is about a half of the total mass
calculated from the components' spectral types of G5III, M dwarf,
A0V and A2V stars, of $7.4 M_{\sun}$.

The \rvs\ from each data set were shifted to the zero point of
$\gamma$. The final orbital solution was calculated from a data
set containing all relative velocities with the data weighted in
the ratio Balona:Collier Cameron:Murdoch:this work
0.02:0.08:0.7:1.0 in accordance with the calculated rms scatter of
each data set solution.
The period of a solution from these combined data was $61.4370 \pm
0.0012$ days with an rms residual of 380 m\,\ps. This orbital
solution is shown in Table \ref{tab:HD101379}. Thus this orbital
solution gives a higher precision for the period.

\subsection{HD\,124425}

The solar neighbourhood star HD\,124425 (HR\,5317) is a short
period spectroscopic binary ($P \approx 2.69$ days). It was found
to have a variation in its radial velocity by A.M.\ Brayton in
early 1920 using Mt Wilson 60-inch telescope spectra
\citep{jcd:1921}.

The Hipparcos satellite photometry measured a magnitude of $H_{\rm p} = 5\mag9996 \pm
0\mag0006$ for this system and indicated that no variability was
detected. The $V$ magnitude from the TYCHO photometry is $5\mag948
\pm 0\mag004$.



HD\,124425 was later observed by \citet{mm+tm:1987} 
with the CORAVEL radial-velocity scanner at Haute-Provence
Observatory between 1980 and 1982. They analysed 16 radial
velocities with an assumed circular orbit and a fixed orbital
period. 

\begin{figure}
    \centering
               \subfigure[Phase plot of HD\,124425 \rvs.]{\includegraphics[width=9.25cm]{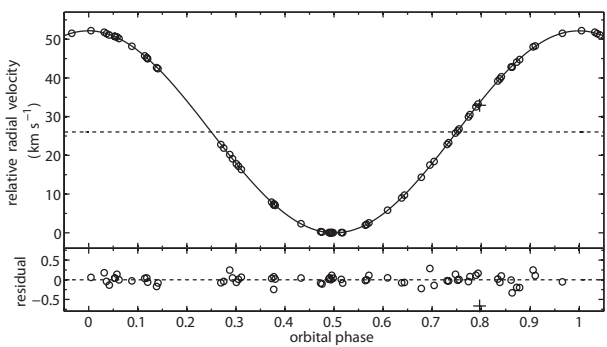}}\\
               \subfigure[Measured \rvs\ of HD\,124425 versus Julian date.]{\includegraphics[width=9.25cm]{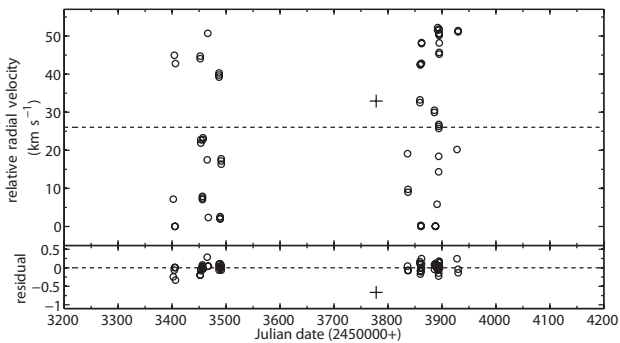}}\\
    \caption[Radial velocity curves of HD\,124425]{Radial velocities of HD\,124425 measured from Hercules spectra.
    The second panels show the residuals from the fit with all free parameters as in table \ref{tab:HD124425}.}
    \label{fig:HD124425}
\end{figure}

In this research, 64 Hercules spectra of HD\,124425 were obtained.
Table \ref{tab:HD124425} is the orbital solution from these
velocities, analysed using SpecBin. A very small and marginally
significant eccentricity was obtained ($e = 0.00260 \pm 0.00099$).
These radial velocities are plotted versus orbital phase in Fig.\
\ref{fig:HD124425}(a) and versus Julian Date in Fig.\
\ref{fig:HD124425}(b). The time of zero mean longitude in Table
\ref{tab:HD124425} ($T_0 = 245\,3800.308\,08$) was used for the
zero point of phase. The lower panels of these figures show the
residuals from the fitted solution. The rms residual velocity is
121 m\,\ps.

\begin{table*}
    \caption[Orbital solutions of HD\,124425 from Hercules data.]{The new orbital parameters of SB1
    HD\,124425 (F6IV) analysed from Hercules spectra.}
    \label{tab:HD124425}
    \begin{center}
        \begin{tabular}{ l l l}
            \hline
   \textbf{Parameter}
                    &   \multicolumn{2}{c}{\textbf{Hercules solutions}} \\
                                                            & \small{with all free parameters}
                                                            & \small{with fixed $P$}       \\
            \hline
   \textit{K} (km\,\ps)          & 26.094 $\pm$ 0.023       & 26.107 $\pm$ 0.023    \\
   \textit{e}                    & 0.002\,60 $\pm$ 0.000\,99
                                                            & $0.002\,70 \pm 0.000\,94$ \\
   \textit{$\omega$} ($^\circ$)  & 294 $\pm$ 26             & 306 $\pm$ 21  \\
   \textit{$T_0$} (HJD)          & 245\,3800.308\,08        &   245\,3800.307\,52   \\
                                 & $\pm$ 0.000\,47          &   $\pm$ 0.000\,40 \\
   \textit{P} (days)             & 2.697\,0329              & 2.697\,0220 (fixed) \\
                                 &   $\pm$ 0.000\,0050      &  --     \\
   \textit{$\gamma_{\mathrm{rel}}$} (km\,\ps) & $26.040 \pm 0.015$
                                                            & 26.035 $\pm$ 0.016    \\
   \textit{$\gamma$} (km\,\ps)   & 18.684 $\pm$ 0.054       &   18.679 $\pm$  0.055 \\
   \textit{$\#_{obs}$}           & 64                       & 64                \\
   \textit{$\#_{rej}$}           & 1                        & 1                 \\
   \textit{$\sigma$} (km\,\ps)   & 0.121                    & 0.126         \\
   \textit{$a_1\sin i$} ($\t 10^7$ km)  & $0.096774 \pm 0.000086$ & $0.096822 \pm 0.000085$      \\
   \textit{$f(M)$} ($M_{\odot}$) & $0.004965 \pm 0.000013$  & $0.004972 \pm 0.000013$ \\
        \hline
        \end{tabular}
    \end{center}
\end{table*}

In order to get a higher precision for the orbital period, the
number of orbital cycles was calculated to be $N_c = 11\,515$ from
the difference in $T_0$ of the recalculated solution of \citet{jcd:1921} 
and the above Hercules solution. This gave $\bar{P}$ = 2.697\,0220
$\pm$ 0.000\,0024 or a precision of 0.2 second. This is half of
the error in the orbital period of \citet{mm+tm:1987}, 
which they calculated in the same way. The orbital solution was
recalculated with this fixed $\bar{P}$ and is also shown in Table
\ref{tab:HD124425}.

\subsection{HD\,136905}

The binary star HD\,136905 (GX Librae) is an ellipsoidal variable
with an active K giant. The star was first reported to have H and
K emission by \citet{wpb+djm:1973}. 
They classified the star as a K1III+F system with the remark that
the emission was uncertain. \citet{ewbetal:1982} 
concluded from their spectroscopic and $UBV$ photometric
observations that this system is an RS CVn-type binary of
spectral type K0III-IV and with a moderate emission at H and K.

\citet{fcfetal:1985} 
reported moderate strength Ca\,II H and K and ultraviolet emission
features and a strong H$\alpha$ absorption.  They suggested that
HD\,136905 is an ellipsoidal variable, as their light curve showed
a frequency of twice the orbital frequency. They also found that
the light curve has a variable amplitude and suggested that this
could be the result of spot activity.

\begin{figure}
    \centering
     \subfigure[]{\includegraphics[width=8.5cm]{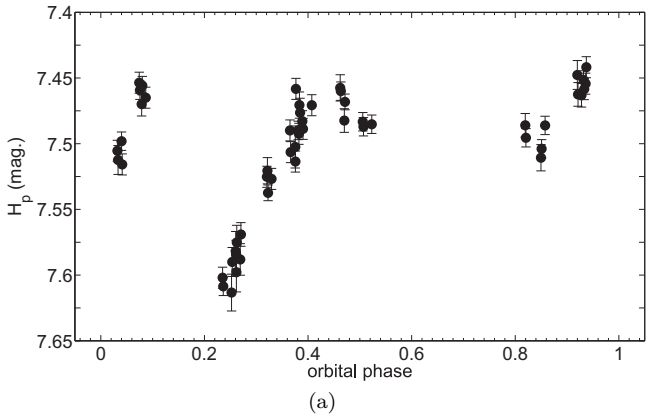}}\\
        \subfigure[]{\includegraphics[width=8.5cm]{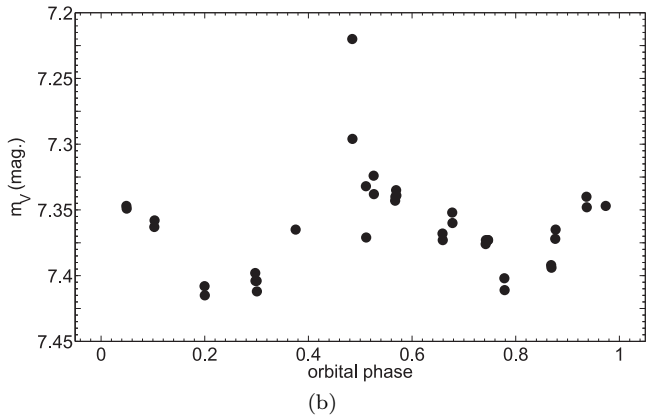}}
        \caption[Light curves of HD\,136905]{The light curves of the active chromosphere variable HD\,136905
        (a) from the Hipparcos satellite and (b) from Mt John Observatory. The orbital phase was calculated from the
        Hercules eccentric solution (Table \ref{tab:HD136905}).}
    \label{fig:HD136905HIP}
\end{figure}

The Hipparcos photometry indicated that HD\,136905 has an
11.12-day periodic variability of semi-amplitude about 0\mag1.
These variations are synchronous with the orbital period and
therefore suggest the spot wave of a tidally locked co-rotating
active chromosphere star.  The $H_p$ magnitude (observed 1989--93)
as well as the $V$ magnitude observed from Mt John photoelectric
photometry (observed 2004--07) are plotted in Fig.\
\ref{fig:HD136905HIP}. The shapes of the curves differ somewhat,
and the semi-amplitude of the Mt John data is about 0\mag05, which
is comparable with the result of \citet{abketal:1995}. 


Previous studies of the orbit of this system were by
\citet{fcfetal:1985}, by \citet{lab:1987} and by
\citet{abketal:1995}. All these authors obtained circular orbits
from data of only moderate precision.

\begin{figure}
    \centering
        \subfigure[Phase plot of HD\,136905 \rvs.]{\includegraphics[width=9.25cm]{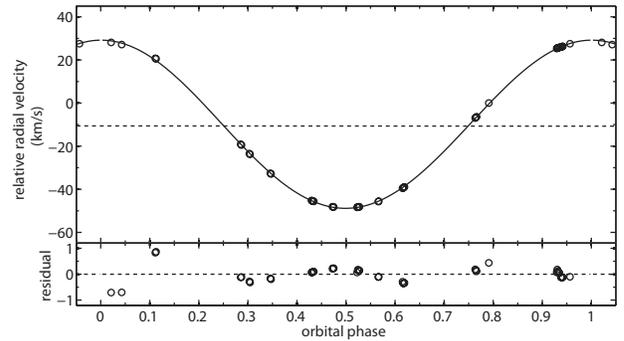}}\\
        \subfigure[Measured \rvs\ of HD\,136905 versus Julian date.]{\includegraphics[width=9.25cm]{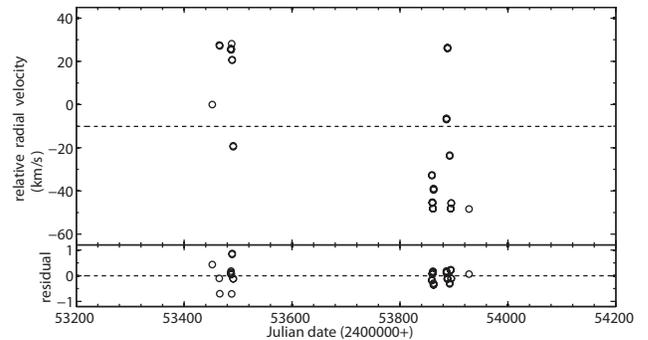}}
    \caption[Radial velocity curves of HD\,136905]{Radial velocities of HD\,136905 measured from the Hercules spectra.
    The second panels show the residuals from that fit.}
    \label{fig:HD136905}
\end{figure}

Our observations with Hercules resulted in 42 spectra of
HD\,136905 with a resolving power of 41\,000.  The orbital
solution is shown in Table \ref{tab:HD136905}. The plot of the
radial-velocity curve from this solution is shown in Fig.\
\ref{fig:HD136905} versus (a) orbital phase and (b) Julian date.
The orbital phase is calculated from the period of an eccentric
solution and we define the time of zero phase as $T_0 =
\mathrm{JD}\,245\,3599.1743 \pm 0.0055$. The rms scatter is 326
m\,\ps, which is high, as can be expected for an ellipsoidal RS
CVn-type binary system.

\begin{table*}
    \caption[Orbital solutions of HD\,136905 from Hercules data]{The orbital parameters of
    SB1 HD\,136905 (K1III) analysed from Hercules \rvs.}
    \label{tab:HD136905}
    \begin{center}
        \begin{tabular}{ l l l l}
            \hline
   \textbf{Parameter} & \multicolumn{3}{c}{\textbf{Hercules solutions}} \\
                                    & \small{eccentric orbit} & \small{with $e = 0$}
                                    & \small{with $P = \bar{P}$}\\
            \hline
   \textit{K} (km\,\ps) & 39.029 $\pm$ 0.078        & 38.989 $\pm$ 0.082
                                            & 39.040 $\pm$ 0.077 \\
   \textit{e}                   & 0.0079 $\pm$ 0.0026   & 0 (fixed)
                                            & 0.0081 $\pm$ 0.0019   \\
   \textit{$\omega$} ($^\circ$) & 23 $\pm$ 24   & - & 33 $\pm$ 17   \\
   \textit{$T_0$} (HJD) & 245\,3599.1743    & 245\,3599.1715    & 245\,3599.1717 \\
                                            & $\pm$ 0.0055      & $\pm$ 0.0067      & $\pm$ 0.0049  \\
   \textit{P} (days)        & 11.134\,14            &   11.134\,13  & 11.134396 (fixed) \\
                                            & $\pm$ 0.000\,30   & $\pm$ 0.00029     &   \\
   \textit{$\gamma_\mathrm{rel}$} (km\,\ps) & $-$10.116 $\pm$ 0.065  & $-$10.055 $\pm$ 0.062  & $-$10.095 $\pm$ 0.060   \\
   \textit{$\gamma$} (km\,\ps)  & 61.78 $\pm$ 0.14      & 61.72 $\pm$ 0.13
                                                            & 61.76 $\pm$ 0.13 \\
   \textit{$\#_{obs}$}                  & 42                & 42            & 42        \\
   \textit{$\#_{rej}$}                  & 0                 & 0             & 0         \\
   \textit{$\sigma$} (km\,\ps)  & 0.326         & 0.385     & 0.325 \\
   \textit{$a_1\sin i$} ($\t 10^7$ km)  & $0.5975 \pm 0.0012$ & $0.5969 \pm 0.0012$ & $0.5977 \pm 0.0012$      \\
   \textit{$f(M)$} ($M_{\odot}$) & $0.06858 \pm 0.00041$ & $0.06837 \pm 0.00043$ & $0.06864 \pm 0.00041$ \\
        \hline
        \end{tabular}
    \end{center}
\end{table*}

The long time span between the historical data-set and the recent
Hercules spectra allows us to determine the orbital period of this
system with a higher precision. The orbital period calculated from
the $T_0$ values of the eccentric solutions of
\citet{abketal:1995} and from Hercules data   is $\bar{P} =
11.134\,396 \pm 0.000\,031$ days (an uncertainty of $\pm 2.7$ s).
However the orbital solution using this more precise period
differs negligibly from that derived from  Hercules data alone, as
seen from Table \ref{tab:HD136905}.

\subsection{HD\,194215}

HD\,194215 (HR\,7801) is a single-lined spectroscopic binary
system for which the primary has been variously classified as K0
(in the HD Catalogue), K3V \citep{wb:1962} and G8II/III
\citep{nh:1982}. The Hipparcos parallax of $6.50 \pm 0.78$ mas
gives $M_V = -0.10 \pm 0.26$, so the primary star is clearly a
giant. \citet{wb+pmm:1958} noted that the \rv\ was variable.
However, there is no evidence of a photometric variation from the
Hipparcos photometry, the Hipparcos magnitude being $H_{\rm p} =
6.0227 \pm 0.0006$.




\citet{bwbetal:1970} 
measured 28 radial velocities of this system and analysed its
orbit. The solution found is an eccentric orbit with \mbox{$e =
0.0687 \pm 0.0138$} and
\mbox{$P = 377.60 \pm 0.25$ days}. 
The rms scatter of their measurement is given as 2.2594 km\,\ps
(so many digits seem unnecessary!). Their data were reanalysed
using SpecBin and it was found that the orbital solution was
significantly different from the published solution, especially
for the orbital period. 
It is possible that the solution or data reported by
\citet{bwbetal:1970} contains a typographical or computational
error, especially as their quoted value of $T$ lies far outside
the time range of the observations.

\begin{figure}
    \centering
       \subfigure[Phase plot of HD\,194215 \rvs.]{\includegraphics[width=9.25cm]{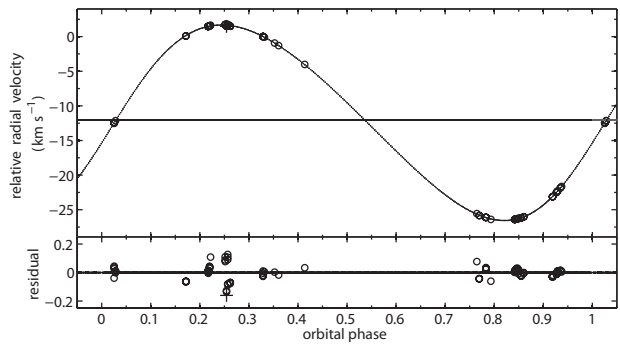}}\\
        \subfigure[Measured \rvs\ of HD\,194215 versus Julian date.]{\includegraphics[width=9.25cm]{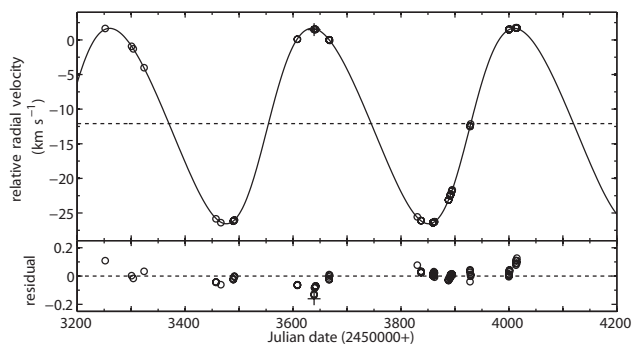}}
    \caption[Radial velocities curves of HD\,194215]{Radial velocities of HD\,194215 measured from the Hercules spectra.
    The plotted curves are calculated from the orbital solution calculated from these \rvs. The second panels show the
    residuals from this fit.}
    \label{fig:HD194215}
\end{figure}

In total, we have obtained 97 Hercules spectra of HD\,194215. An
eccentric orbital solution was calculated from the \rvs\ measured
from these spectra. This solution is shown in Table
\ref{tab:HD194215}. The \rvs\ are shown in Fig.\
\ref{fig:HD194215} plotted versus (a) orbital phase and (b) Julian
date. This eccentric solution has an rms scatter of the fit of 47
m\,\ps.  The time of periastron passage is $T =
\mathrm{JD}\,245\,3918.52 \pm 0.65$. For this star, we have chosen
the zero point for phase to be the time of periastron.
\begin{table}
    \caption[Orbital solutions of HD\,194215 from Hercules data.]{The orbital parameters of
    SB1 HD\ 194215 (G8II/III) analysed from Hercules \rvs.}
    \label{tab:HD194215}
    \begin{center}
        \begin{tabular}{ l l }
            \hline
   \textbf{Parameter}   &  \textbf{Hercules solutions} \\
                            & \small{all free parameters} \\
            \hline
   \textit{K} (km\,\ps) & 14.1155 $\pm$ 0.0056          \\
   \textit{e}           & 0.123\,29 $\pm$ 0.000\,78 \\
   \textit{$\omega$} ($^\circ$) & 258.14 $\pm$ 0.77 \\
   \textit{$T_0$} (HJD) & 245\,3649.711 $\pm$ 0.074 \\
   \textit{P} (days)        & 374.88 $\pm$ 0.18                 \\
   \textit{$\gamma_{\mathrm{rel}}$} (km\,\ps)   & $-12.086 \pm 0.022$ \\
   \textit{$\gamma$} (km\,\ps)  & $-$8.14 $\pm$ 0.14    \\
   \textit{$\#_{obs}$}                  & 97                \\
   \textit{$\#_{rej}$}                  & 1                 \\
   \textit{$\sigma$} (km\,\ps)  & 0.047         \\
   \textit{$a_1\sin i$} ($\t 10^7$ km)  & $7.2210 \pm 0.0063$ \\
   \textit{$f(M)$} ($M_{\odot}$) & $0.10676 \pm 0.00018$ \\
        \hline
        \end{tabular}
    \end{center}
\end{table}


%

\section{The reality of small detected eccentricities}
\label{nonkepvel}

In this section we analyse the reality of the small eccentricities
found for the 12 stars discussed in this paper and by
\citet{sketal:2011}. The tests used are those described by
\citet{ll:2005} and also by \citet{lbl+mas:1971}.

The first Lucy test calculates the probability ($p_1$)  that the
star really has a circular orbit, but by chance the data show a
small eccentricity, shown by a second harmonic in the velocity
curve. If $p_1 < 0.05$, the eccentricity found is deemed to be
significant.

However, the possibility remains that the second harmonic may be
caused by a non-keplerian perturbation.   If the data indicate a
keplerian eccentric orbit, then the third harmonic must have the
amplitude and phase which are consistent with the observed second
harmonic. As pointed out by \citet{ll:2005}, the components of the
keplerian third harmonic are $C_3\cos 3L = \frac{9}{8}Ke^2\cos
2\omega \cos 3L$ and $S_3\sin 3L = \frac{9}{8}Ke^2\sin 2\omega
\sin 3L$, where $L=2\pi(t-T)/P + \omega$ is the mean longitude.
These compare with the generally much larger keplerian second
harmonic components of $C_2\cos 2L = Ke\cos\omega\cos 2L$ and
$S_2\sin 2L = Ke\sin\omega\sin 2L$.

The second Lucy test is to add a non-keplerian third harmonic
perturbation to the eccentric keplerian orbit $V_{\rm K}$ so that
the observed velocities become $V_{\rm obs} = V_{\rm K} + \Delta
C_3\cos 3L + \Delta S_3 \sin 3L$. The observed estimate of the
amplitude of the non-keplerian perturbation is designated as
$\delta_3 = (\delta C_3, \delta S_3)$. The probability $p_2$
measures the probability that the data indicate a perturbation by
chance, even though a purely keplerian eccentric orbit is being
followed. If $p_2 < 0.05$, then the perturbation is deemed to be
real; a larger value indicates a keplerian orbit.

\citet{ll:2005} has proposed two further tests. The third is to
find the probability ($q_1$) that no keplerian third harmonic has
been detected, and the fourth is to find the probability ($q_2$)
that no third harmonic, whether keplerian or from a perturbation,
has been detected.  Lucy defines the parameter $p_3$ as the larger
of $q_1$ and $q_2$ and uses the parameters $p_2, p_3$ to divide
the stars into four classes:
\begin{description}
\item[\bf class A] $p_2 > 0.05; p_3 < 0.05$ An unperturbed
keplerian eccentric orbit is strongly supported.

\item[\bf class B] $p_2 < 0.05; p_3 < 0.05$ An eccentric orbit has
some support. However, a perturbation is detected, so the orbital
elements are in doubt.

\item[\bf class C] $p_2 > 0.05; p_3 > 0.05$ The eccentricity is
not strongly supported. Although no perturbation in the third
harmonic is detected, no keplerian third harmonic is detected
either. The second harmonic may arise from a perturbation or a
small eccentricity.

\item[\bf class D] $p_2 < 0.05; p_3 > 0.05$ An eccentric keplerian
orbit is in doubt. The keplerian third harmonic is not detected,
but a third harmonic arising from a perturbation is detected, as
shown by its amplitude and phase.
\end{description}

The results of all these tests are shown in Table \ref{lucy}. In
this table the probabilities in column 2 are defined by
\citet{ll:2005}. In column 3, ($C_2, S_2$) are the components of
the second harmonic derived from a keplerian fit. In column 4,
($\delta C_3, \delta S_3$) are the amplitudes of the orthogonal
components of the non-keplerian perturbation obtained from our
Hercules data. Column 5 gives the expected keplerian third
harmonic amplitude calculated from our orbital elements. The
eccentricity (possibly spurious) from a fitted keplerian orbit is
in column 6.  The last column is self-explanatory.

In summary, the results are as follows: four stars (HD\,22905,
HD\,38099, HD\,85622 and HD\,101379) show $p_1 > 0.05$ so for
these there is no reason to invoke a non-circular orbit. We
conclude that the small eccentricities reported in the earlier
literature are spurious. Nevertheless, for completeness, we have
applied all the Lucy tests to all the stars, as even for these
stars, a perturbation third harmonic could in principle be
detected.

\begin{table*}
    \caption[Results of the Lucy tests.]{The results of harmonic analysis of the velocity data using tests by Lucy.}
    \label{lucy}
    \begin{center}
        \begin{tabular}{ l l c c c c c }
            \hline
   \textbf{Star}   &  \textbf{Probabilities} & $(C_2,S_2)$  & $(\delta C_3, \delta S_3$) &
   $(C_3, S_3)$ & $e$ & \textbf{Lucy} \\
   & $p_1, p_2, q_1, q_2, p_3$  & ms$^{-1}$ & ms$^{-1}$     & ms$^{-1}$ &     & \textbf{class} \\
\hline

HD\,352  & $p_1 =  9.953e-29$ & $C_2 = -163.90$ & $\delta C_3 = -11.29 \pm 15.07$ & $C_3 = -6.71$ & $0.023$ & D \\
         & $p_2 =  1.555e-04$ & $S_2 = 693.36$  & $\delta S_3 = -66.04 \pm 15.67$ & $S_3 = -13.76$ & $\pm 0.0006$ & \\
         & $q_1 =  5.444e-01$ & \\
         & $q_2 =  4.233e-06$ & \\
         & $p_3 =  5.444e-01$ & \\
         \\
HD\,9053 & $p1 =  1.348e-44$  & $C_2 = -179.29$ & $\delta C_3 =  -25.59 \pm 13.41$ & $C_3 =    1.09$ & 0.0301 & C \\
         & $p_2 = 1.048e-01$  & $S_2 = 323.45$  & $\delta S_3 =    1.03 \pm 19.59$ & $S_3 =  -16.65$ & $\pm 0.0012$ \\
         & $q_1 = 6.151e-01$ \\
         & $q_2 = 1.936e-01$ \\
         & $p_3 = 6.151e-01$ \\
         \\
HD\,22905 & $p_1 =  1.690e-01$ & $C_2 = -9.51$  & $\delta C_3 =   47.71 \pm 12.85$ & $C_3 = -0.04$ & 0.0010       & D \\
          & $p_2 =  6.408e-04$ & $S_2 = 272.40$ & $\delta S_3 =   18.63 \pm 12.55$ & $S_3 =  0.05$ & $\pm 0.0014$ \\
          & $q_1 =  1.000$     & \\
          & $q_2 =  6.428e-04$ & \\
          & $p_3 =  1.000$     & \\
          \\
HD\,30021 & $p_1 =  1.622e-03$ & $C_2 = 33.39$  & $\delta C_3 =  -15.47 \pm 10.76$  & $C_3 = -0.05$ &  0.0015 & C \\
          & $p_2 =  3.353e-01$ & $S_2 = 33.39$ & $\delta S_3 =   -3.91 \pm 12.95$  & $S_3 =  0.08$ & $\pm 0.0015$ \\
          & $q_1 =  1.000$     & \\
          & $q_2 = 3.338e-01$  & \\
          & $p_3 = 1.000$      & \\
          \\
HD\,38099 & $p_1 =  5.228e-01$ & $C_2 = -59.46$  & $\delta C_3 =  -33.86 \pm 48.76$  & $C_3 =   -0.50$ & 0.0076 & C \\
          & $p_2 =  6.531e-01$ & $S_2 = 14.83$  & $\delta S_3 =  -46.73 \pm 55.24$  & $S_3 =   -0.29$ & $\pm 0.0067$ \\
          & $q_1 =  9.999e-01$ & \\
          & $q_2 =  6.481e-01$ & \\
          & $p_3 =  9.999e-01$ & \\
          \\
HD\,50337 & $p_1 =  2.073e-31$ & $C_2 = 6.29$  & $\delta C_3 =   38.11 \pm 5.20$ & $C_3 =   -0.88$ & 0.0059   & D \\
          & $p_2 =  1.567e-10$ & $S_2 = 105.86$ & $\delta S_3 =   12.44 \pm 5.12$ & $S_3 =    0.30$ & $\pm 0.0003$ \\
          & $q_1 =  9.839e-01$ & \\
          & $q_2 =  2.979e-10$ & \\
          & $p_3 =  9.839e-01$ & \\
          \\
HD\,77258 & $p_1 =  4.260e-04$ & $C_2 = -4.61$  & $\delta C_3 =   -6.82 \pm 2.54$ & $C_3 =   -0.01$ & 0.0009 & D \\
          & $p_2 =  2.927e-02$ & $S_2 = 16.08$  & $\delta S_3 =   -0.32 \pm 2.94$ & $S_3 =   -0.01$ & $\pm 0.0002$ \\
          & $q_1 =  1.000$     & \\
          & $q_2 =  2.885e-02$ & \\
          & $p_3 =  1.000$     & \\
          \\
HD\,85622 & $p_1 =  1.517e-01$ & $C_2 = -33.50$  & $\delta C_3 =   16.91 \pm 9.66$ & $C_3 =    0.10$ & 0.0026 & C \\
          & $p_2 =  9.079e-02$ & $S_2 = -4.71$  & $\delta S_3 =   -9.98 \pm 8.68$ & $S_3 =    0.02$ & $\pm 0.0016$ \\
          & $q_1 =  9.999e-01$ & \\
          & $q_2 =  8.949e-02$ & \\
          & $p_3 =  9.999e-01$ & \\
          \\
HD\,101379& $p_1 =  4.467e-01$ & $C_2 = -84.38$  & $\delta C_3 =  -72.75 \pm 72.68$ & $C_3 =   -0.94$ & 0.012 & D \\
          & $p_2 =  4.998e-06$ & $S_2 = -129.94$  & $\delta S_3 = -360.61 \pm 67.20$ & $S_3 =    2.02$ & $\pm 0.010$ \\
          & $q_1 =  9.994e-01$ \\
          & $q_2 =  5.587e-06$ \\
          & $p_3 =  9.994e-01$ \\
          \\
HD\,124425& $p_1 =  2.570e-02$ & $C_2 = 27.59$  & $\delta C_3 =   14.78 \pm 20.54$ & $C_3 =   -0.13$ & 0.0026 & C \\
          & $p_2 =  7.198e-01$ & $S_2 = -0.91$ & $\delta S_3 =    7.89 \pm 21.01$ & $S_3 =   -0.15$ & $\pm 0.0010$ \\
          & $q_1 =  1.000$     & \\
          & $q_2 =  7.249e-01$ & \\
          & $p_3 =  1.000$     & \\
          \\
\hline
        \end{tabular}
    \end{center}
\end{table*}

\addtocounter{table}{-1}
\begin{table*}
    \caption[Results of the Lucy tests (cont.).]{The results of harmonic analysis
    of the velocity data using tests by Lucy (cont.).}
    \label{lucy}
    \begin{center}
        \begin{tabular}{ l l c c c c c }
            \hline
   \textbf{Star}   &  \textbf{Probabilities} & $(C_2,S_2)$  & $(\delta C_3, \delta S_3$) &
   $(C_3, S_3)$ & $e$ & \textbf{Lucy} \\
   & $p_1, p_2, q_1, q_2, p_3$  & ms$^{-1}$ & ms$^{-1}$     & ms$^{-1}$ &     & \textbf{class} \\
\hline

HD\,136905& $p_1 =  1.003e-03$ & $C_2 = 283.82$  & $\delta C_3 = -319.42 \pm 58.30$ & $C_3 =    1.86$ & 0.0079 & D \\
          & $p_2 =  9.895e-07$ & $S_2 = 120.47$  & $\delta S_3 =   39.17 \pm 46.91$ & $S_3 =    1.98$ & $\pm 0.0026$ \\
          & $q_1 =  9.990e-01$ & \\
          & $q_2 =  1.026e-06$ & \\
          & $p_3 =  9.990e-01$ & \\
          \\
HD\,194215& $p_1 = 1.422e-113$ & $C_2 = -357.67$  & $\delta C_3 =  0.58 \pm 7.03$ & $C_3 = -220.98$ & 0.1233 & A \\
          & $p_2 =  9.850e-01$ & $S_2 = -1703.15$ & $\delta S_3 = -0.93 \pm 6.43$ & $S_3 =   97.18$ & $\pm 0.0008$ \\
          & $q_1 =  3.326e-56$ \\
          & $q_2 =  4.754e-56$ \\
          & $p_3 =  4.754e-56$ \\
          \\

\hline
        \end{tabular}
    \end{center}
\end{table*}

We find one of the 12 stars, HD\,194215, to be in class A. No
perturbation is detected for this star, and a fairly large
non-zero eccentricity ($e=0.1233 \pm 0.0001$) is clearly present,
even though \citet{bwbetal:1970} had earlier reported a smaller
value ($e = 0.06 \pm 0.01$).

We find none of our stars to be in Lucy class B. Five are in class
C, where neither a keplerian third harmonic nor a perturbation
third harmonic is detected, and six are in class D, where no
keplerian third harmonic is detected, but a perturbation third
harmonic is clearly seen. We can conclude that all the class D
stars are affected by non-keplerian perturbations and probably
have circular orbits. It is also possible that the class C stars
also have circular orbits. We note that the keplerian third
harmonics are nearly all predicted to have very small amplitudes,
of the order of 1 ms$^{-1}$ in most cases, except for HD\,194215,
where it is a large 241 ms$^{-1}$. It is not surprising therefore
that the keplerian third harmonics have been undetectable (with
one exception) using our data with a typical precision per
observation of about 10 ms$^{-1}$ in the best cases. On the other
hand, the observed non-keplerian perturbation amplitudes of the
third harmonic are occasionally quite large; for example 67
ms$^{-1}$ for HD\,352, 51 ms$^{-1}$ for HD\,22905, 322 ms$^{-1}$
for HD\,136905 and 368 ms$^{-1}$ for HD\,101379. Details are in
Table \ref{lucy}. In all these cases, our data are readily able to
detect such third harmonic perturbations.

For the Class D stars, as these are presumed to have circular
orbits, the second harmonic must then arise entirely from
non-keplerian effects, and this is commonly several hundred
ms$^{-1}$. For the class C stars, the second harmonic may contain
both keplerian and non-keplerian velocities.

\section{Possible causes of non-keplerian velocity perturbations}

It is interesting to consider the possible causes for
non-keplerian perturbations. The most likely cause, originally
suggested by \citet{os:1950}, is starspots on a rotating star,
which render the surface brightness of the visible hemisphere
non-uniform and lead to rotational modulation of the observed
radial velocity. Even for the Sun, which has an equatorial
rotational velocity of about 2 km\,s$^{-1}$, a small number of
starspots can easily perturb the mean velocity that would be
observed for the visible hemisphere by several ms$^{-1}$. Given
that our stars are giants and typically ten times larger than the
Sun in radius, this magnifies the susceptibility of spots to
perturb the observed radial velocity (for a given rotation
period).  Moreover, several stars in our list are known to be
chromospherically active (HD\,9053, HD\,50337, HD\,136905,
HD\,101379), so these probably have starspots.

We also note that another star ($\zeta$ TrA) that we had earlier
analysed, also showed a small eccentricity ($e = 0.0140 \pm
0.0002$) \citep{jsetal:2004} from our Hercules data. This star was
subsequently shown by \citet{ll:2005} to be in class D, that is
with no significant keplerian third harmonic but a significant
perturbation third harmonic. We subsequently did tests with the
Wilson-Devinney code \citep{rew+ejd:1971} to model the effect of
starspots on the observed radial velocity curve assuming a truly
circular orbit. It was a simple matter to obtain spurious
eccentricities comparable to those observed for $\zeta$ TrA by
assuming a co-rotating model with a group of spots at about the
same longitude and near the equator covering 12 per cent of the
surface and with $\Delta T \sim 1000$ K \citep{sketal:2006}, or
alternatively by adopting a model with two groups of spots each
covering 8 per cent of the star's surface area. For the class C
and D stars in this paper and in \citet{sketal:2011} the
eccentricities found range up to $e = 0.03$ for HD\,9053, but the
values are mainly less than $e = 0.01$, so we presume that spots
can rather easily perturb the radial-velocity data to produce
spurious eccentricities of this order.

Another possibility is that non-spherical tidally distorted stars
with significant gravity darkening can also, as a result of their
variation in surface brightness over the visible part of the
surface, contribute to velocity perturbations, as discussed in
some detail by \citet{tes:1941a}. We consider this less likely for
our sample of stars, because they mainly have orbital periods of
several months to a year. Only HD\,124425 has an orbital period of
less than 10 days and is therefore likely to be more affected by
tidal distortion caused by the secondary star. But even here, this
is a Lucy class C star for which no perturbation third harmonic
was detected. However, two stars we have observed are known to
show ellipsoidal light variations, notably HD\,136905, and
possibly also HD\,38099.

A third possibility is that the underlying secondary star spectrum
has distorted our measurements of the radial velocity of the
primary of our SB1 systems. We have not attempted to model this
effect, but note that our stars all have giant, or even
supergiant, primaries. It is likely that a common situation is a G
or K giant primary with an M dwarf secondary, possibly as much as
ten magnitudes fainter, given the large preponderance of M dwarfs
in stellar statistics. Nevertheless three of our stars are thought
to have F dwarf secondaries (HD\,352, HD\,22905, HD\,136905) and
two to have A dwarf secondaries (HD\,50337, HD\,77258). We
acknowledge that if the visual luminosity of the secondary just
happens to lie at more than two but less than four magnitudes
below the primary, then it might have a disturbing effect on the
primary star's velocity measurements. We expect that secondaries
at less than two magnitudes below the primary would have been
detected and the system would have been classified as SB2, while
those at more than four magnitudes, which is probably the case for
the great majority of our stars, would have been too faint to
cause a problem, as the secondary star's spectrum would be
comparable to or less than the photon noise in the recorded
spectrum. The secondary spectrum in principle has the possibility
of perturbing the measured velocities for almost any method of
measuring Doppler shifts, whether by cross-correlation or some
other technique. The perturbation would depend not only on the
magnitude difference of the two stars, but also on the spectral
type and radial velocity difference.

Although any of these processes may be the source of non-keplerian
velocities, starspots and chromospheric activity remain the most
likely explanation, especially in view of the fact noted that the
most chromospherically active stars in our sample have the largest
velocity perturbations.

\section{Conclusion}

We have analysed the orbits of six single-lined spectroscopic
binary systems with nearly circular orbits, using high precision
radial-velocity data from the Hercules fibre-fed vacuum
spectrograph at Mt John Observatory. We have paid especial
attention to obtaining precise (i.e. small random errors) values
of the eccentricity by fitting keplerian orbits by the least
squares method to the data points.

We have combined the results of the orbital analyses in this paper
with six further SB1 orbits analysed by \citet{sketal:2011}.

Our results show that only for one star (HD\,194215) is there a
clear and unequivocal orbital eccentricity detected ($e=0.1233 \pm
0.0001$). For four stars (HD\,22905, HD\,38099, HD\,85622 and
HD\,101379) our analysis of the data show that a circular orbit is
the preferred solution. For six stars we find clear evidence of
small systematic perturbations in the measured radial velocities
which have produced spurious values of the eccentricity. These
spurious eccentricities are mainly below $e = 0.01$, but in one
case it is as high as $e = 0.03$ (HD\,9053) and in another case it
is $e = 0.02$ (HD\,352). Our analysis also gives quantitative
estimates of the perturbation in second and third harmonics;
values as small as 10 m\,\ps\ can be detected, and values over 300
m\,\ps\ can occur in moderately active stars.

Our overall conclusion is therefore that no matter how precise or
numerous the radial-velocity data may be, conventional methods of
velocity determination, such as by cross-correlation, lead to
unavoidable systematic errors in the velocities, which are not
random but depend on the phase, and which may amount to a few tens
of ms$^{-1}$ to a few hundreds of ms$^{-1}$ in some cases.
Therefore it is likely that the great majority of late-type giant
spectroscopic binary orbital eccentricities cannot be trusted to
an accuracy of better than a few times 0.01 in $e$. The only way
to overcome this problem would be much more careful modelling of
effects such as starspots, taking into account their distribution
on the stellar surface and effect on line profiles through Doppler
imaging. Our analysis measures the non-keplerian harmonics of
circularized orbits at the 10 ms$^{-1}$ level for the first time,
and demonstrates that circularization even to better than $e=0.01$
has been achieved for many of the stars in our sample.

This project had at its inception the goal of investigating very
small reported eccentricities ($e \leq 0.05$)  using the best high
quality spectra from a high-dispersion vacuum fibre-fed \'echelle
spectrograph, and possibly further to investigate whether complete
circularization had been achieved at the $e=0.01$ level. Instead
we have found that unavoidable systematic effects have hampered
that line of investigation, but nevertheless this has led to new
results of interest on the ability of spectroscopic data to
determine the orbital eccentricity in binary stars and to quantify
non-keplerian velocity perturbations.

\appendix

\tiny 

\setcounter{table}{10}
\renewcommand{\arraystretch}{0.8}
\begin{center}
\begin{table*}
\caption[Radial velocity of HD\,77258]{Radial velocities of
HD\,77258 relative to the image recorded at JD2453741.1562.}
\label{rv:HD77258}

\begin{tabular}{r r r r r r}

\hline
\textbf{HJD } & \textbf{$V_\mathrm{rad}$} & \textbf{HJD } & \textbf{$V_\mathrm{rad}$} & \textbf{HJD } & \textbf{$V_\mathrm{rad}$} \\
(2400000+) & (km\,\ps) & (2400000+) & (km\,\ps) & (2400000+) & (km\,\ps) \\
\hline

53323.1328 & 35.627 & 53453.0742 & 9.326  & 53488.8359 & 29.199 \\
53377.0820 & 6.599 & 53456.8085 & 15.153 & 53488.8398 & 29.202 \\
53402.0781 & 37.863 & 53456.8125 & 15.151 & 53490.8632 & 26.250 \\
53403.0703 & 37.915 & 53456.8164 & 15.183 & 53490.8671 & 26.241 \\
53403.9609 & 37.615 & 53456.8203 & 15.180 & 53490.8710 & 26.238 \\
\\
53404.0234 & 37.691 & 53464.9531 & 28.236 & 53490.8750 & 26.231 \\
53405.0429 & 37.637 & 53465.8750 & 29.512 & 53490.8789 & 26.223 \\
53406.0898 & 37.304 & 53466.8710 & 30.847 & 53666.1562 & $-$0.491  \\
53424.9179 & 12.859 & 53486.8515 & 31.786 & 53666.1601 & $-$0.484 \\
53425.0312 & 12.695 & 53486.8554 & 31.782 & 53666.1640 & $-$0.478 \\
\\
53451.9726 & 7.615 & 53486.8593 & 31.777 & 53689.1718 & 30.698 \\
53451.9765 & 7.625 & 53486.8632 & 31.770 & 53689.1757 & 30.705 \\
53452.9960 & 9.111 & 53488.8320 & 29.220 & 53721.1210 & 13.401 \\
53453.0078 & 9.138 & 53488.8320 & 29.203 & 53721.1250 & 13.380 \\
\\
53741.1484 & 0.002 & 53835.9335 & 28.593 & 53885.8906 & $-$1.333 \\
53741.1523 & $-$0.014 & 53835.9375 & 28.602 & 53885.9023 & $-$1.354 \\
53741.1562 & 0.000 & 53835.9414 & 28.611 & 53887.8242 & $-$0.842 \\
53742.0976 & 0.623 & 53835.9453 & 28.608 & 53887.8281 & $-$0.838 \\
53742.1015 & 0.624 & 53858.9179 & 30.040 & 53887.8320 & $-$0.838  \\
\\
53742.1054 & 0.631 & 53858.9218 & 30.032  & 53887.8359 & $-$0.838 \\
53743.0507 & 1.379 & 53858.9296 & 30.028  & 53890.9257 & 1.042  \\
53743.0585 & 1.379 & 53858.9335 & 30.032 & 53890.9335 & 1.059\\
53744.0859 & 2.350 & 53858.9375 & 30.029 & 53890.9375 & 1.050  \\
53744.0898 & 2.347 & 53860.9375 & 27.166 & 53891.8789 & 1.864\\
\\
53778.0976 & 36.220 & 53860.9375 & 27.156 & 53891.8828 & 1.890 \\
53778.1015 & 36.197 & 53860.9414 & 27.141 & 53891.8867 & 1.899 \\
53778.1093 & 36.201 & 53860.9453 & 27.142 & 53891.8906 & 1.909  \\
53779.1445 & 35.600 & 53861.9375 & 25.635 & 53893.8125 & 3.864  \\
53779.1484 & 35.609 & 53861.9414 & 25.637 & 53893.8125 & 3.873\\
\\
53779.1523 & 35.626 & 53861.9453 & 25.628 & 53893.8203 & 3.872 \\
53829.9492 & 19.223 & 53861.9492 & 25.622 & 53893.8203 & 3.876\\
53829.9531 & 19.233 & 53861.9531 & 25.622 & 53929.8046 & 33.769 \\
53829.9570 & 19.232 & 53861.9531 & 25.615 & 53929.8125 & 33.866 \\
53829.9609 & 19.243 & 53885.8828 & $-$1.333 & & \\
\end{tabular}
\end{table*}
\end{center}

\renewcommand{\arraystretch}{0.8}
\begin{center}
\begin{table*}

\caption[Radial velocity of HD\,85622]{Radial velocities of
HD\,85622 relative to the image recorded at JD2453835.9726.}
\label{rv:HD85622}

\begin{tabular}{r r r r r r}

\hline
\textbf{HJD } & \textbf{$V_\mathrm{rad}$} & \textbf{HJD } & \textbf{$V_\mathrm{rad}$} & \textbf{HJD } & \textbf{$V_\mathrm{rad}$} \\
(2400000+) & (km\,\ps) & (2400000+) & (km\,\ps) & (2400000+) & (km\,\ps) \\
\hline

53323.1523 & $-$20.257 & 53490.9101 & $-$2.112 & 53860.9609 & 1.512 \\
53377.0898 & $-$24.424 & 53490.9140 & $-$2.119 & 53860.9648 & 1.493 \\
53402.0859 & $-$21.613 & 53490.9179 & $-$2.090 & 53860.9687 & 1.493 \\
53403.0781 & $-$21.518 & 53490.9257 & $-$2.098 & 53860.9726 & 1.497 \\
53403.9687 & $-$21.634 & 53692.1718 & $-$24.592& 53861.9804 & 1.484 \\
\\
53404.0351 & $-$24.270 & 53692.1718 & $-$24.592 & 53861.9843 & 1.500 \\
53405.0507 & $-$21.159 & 53741.1640 & $-$19.888 & 53861.9882 & 1.498 \\
53406.0976 & $-$21.044 & 53741.1679 & $-$19.882 & 53861.9882 & 1.502 \\
53424.9375 & $-$17.486 & 53741.1718 & $-$19.875 & 53885.9101 & $-$0.122 \\
53425.0507 & $-$17.526 & 53742.1132 & $-$19.763 & 53885.9179 & $-$0.135 \\
\\
53451.9843 & $-$10.755 & 53742.1210 & $-$19.770 & 53885.9218 & $-$0.128 \\
53451.9921 & $-$10.747 & 53742.1289 & $-$19.760 & 53885.9257 & $-$0.123 \\
53451.9960 & $-$10.738 & 53744.0976 & $-$19.312 & 53887.8437 & $-$0.188 \\
53456.9765 & $-$9.511 & 53744.1054 & $-$19.319 & 53887.8476 & $-$0.201 \\
53456.9804 & $-$9.512 & 53778.1132 & $-$12.055 & 53887.8515 & $-$0.214 \\
\\
53456.9843 & $-$9.500 & 53778.1210 & $-$12.061 & 53887.8554 & $-$0.222 \\
53456.9882 & $-$9.511 & 53778.1289 & $-$11.999 & 53890.9687 & $-$0.736 \\
53465.9062 & $-$7.308 & 53835.9726 & 0.000 & 53890.9765 & $-$0.741 \\
53466.9140 & $-$7.111 & 53835.9804 & 0.008 & 53890.9843 & $-$0.737 \\
53486.8945 & $-$2.785 & 53835.9843 & 0.006 & 53891.8945 & $-$0.787 \\
\\
53486.8984 & $-$2.783 & 53835.9882 & 0.001 & 53891.9023 & $-$0.801 \\
53486.9023 & $-$2.783 & 53835.9960 & $-$0.004 & 53891.9062 & $-$0.807 \\
53486.9023 & $-$2.785 & 53858.9765 & 1.585 & 53893.8437 & $-$1.116 \\
53488.8476 & $-$2.498 & 53858.9843 & 1.591 & 53893.8515 & $-$1.124 \\
53488.8515 & $-$2.493 & 53858.9882 & 1.591 & 53893.8554 & $-$1.139 \\
\\
53488.8554 & $-$2.490 & 53860.9531 & 1.497 & 53893.8554 & $-$1.127 \\
53488.8593 & $-$2.506 & 53860.9570 & 1.516 & 53894.8007 & $-$1.250 \\
\\
53894.8046 & $-$1.243 & 53894.8125 & $-$1.230 & 53929.8242 & $-$8.277 \\
53894.8085 & $-$1.234 & 53928.8359 & $-$8.079 & & \\
\end{tabular}
\end{table*}
\end{center}

\newpage

\renewcommand{\arraystretch}{0.8}
\begin{center}
\begin{table*}

\caption[Radial velocity of HD\,101379]{Absolute radial velocities
of HD\,101379 obtained using HD\,109379 (G5II) as a standard
star.} \label{rv:HD101379}

\begin{tabular}{r r r r r r}

\hline
\textbf{HJD } & \textbf{$V_\mathrm{rad}$} & \textbf{HJD } & \textbf{$V_\mathrm{rad}$} & \textbf{HJD } & \textbf{$V_\mathrm{rad}$} \\
(2400000+) & (km\,\ps) & (2400000+) & (km\,\ps) & (2400000+) & (km\,\ps) \\
\hline

53304.1054 & 14.893 & 53465.0117 & $-$9.838 & 53779.1796 & $-$3.842\\
53322.1640 & 1.617 & 53465.9375 & $-$9.242 & 53779.1875 & $-$3.807\\
53377.1289 & 9.489 & 53466.9531 & $-$8.476 & 53836.0039 & $-$8.143 \\
53402.1015 & $-$10.464 & 53486.9687 & 13.368 & 53859.0000 & 13.280\\
53402.1757 & $-$10.508 & 53486.9765 & 13.367 & 53859.0117 & 13.271\\
\\
53403.0859 & $-$10.000 & 53486.9804 & 13.350 & 53859.0234 & 13.275 \\
53403.1445 & $-$10.035 & 53488.0039 & 13.865 & 53859.9687 & 13.207\\
53404.0156 & $-$9.589 & 53488.0117 & 13.868 & 53859.9804 & 13.203\\
53404.0859 & $-$9.557 & 53488.0234 & 13.867 & 53859.9960 & 13.196\\
53404.1367 & $-$9.470 & 53488.9179 & 14.132 & 53861.0039 & 13.016\\
\\
53405.0585 & $-$8.634 & 53488.9257 & 14.133 & 53861.0078 & 13.031\\
53405.1367 & $-$8.587 & 53488.9335 & 14.111 & 53861.0156 & 13.059 \\
53406.0156 & $-$7.674 & 53490.9335 & 14.324 & 53861.0234 & 13.026\\
53406.1210 & $-$7.669 & 53490.9414 & 14.322 & 53862.0273 & 12.738 \\
53423.9296 & 12.500 & 53490.9492 & 14.354 & 53862.0312 & 12.739\\
\\
53423.9609 & 12.524 & 53742.1679 & 11.287 & 53862.0390 & 12.729\\
53424.9687 & 13.115 & 53744.1367 & 9.605 & 53862.0468 & 12.732\\
53457.0312 & $-$10.842 & 53744.1445 & 9.574 & 53885.9648 & $-$11.554\\
53457.0429 & $-$10.852 & 53779.1757 & $-$3.856 & 53885.9726 & $-$11.570\\
\\
53885.9804 & $-$11.571 & 53891.9570 & $-$11.724 & 53894.8906 & $-$10.226 \\
53885.9882 & $-$11.571 & 53891.9648 & $-$11.721 & 53894.8945 & $-$10.234 \\
53887.8828 & $-$12.165 & 53891.9726 & $-$11.730 & 53927.8320 & 9.865 \\
53887.8906 & $-$12.175 & 53893.8867 & $-$10.885 & 53929.8398 & 7.896 \\
53887.8945 & $-$12.177 & 53893.8906 & $-$10.873 & 53929.8515 & 7.895 \\
\\
53891.0000 & $-$12.066 & 53893.8945 & $-$10.883  &  \\
53891.0117 & $-$12.066 & 53894.8828 & $-$10.259 &  \\

\end{tabular}
\end{table*}
\end{center}

\renewcommand{\arraystretch}{0.8}
\begin{center}
\begin{table*}
\caption[Radial velocity of HD\,124425]{Radial velocities of
HD\,124425 relative to the image recorded at JD2453405.1757.}
\label{rv:HD124425}

\begin{tabular}{r r r r r r}

\hline
\textbf{HJD } & \textbf{$V_\mathrm{rad}$} & \textbf{HJD } & \textbf{$V_\mathrm{rad}$} & \textbf{HJD } & \textbf{$V_\mathrm{rad}$} \\
(2400000+) & (km\,\ps) & (2400000+) & (km\,\ps) & (2400000+) & (km\,\ps) \\
\hline

53402.1640 & 7.174 & 53487.0234 & 40.302    &  53861.9648 & 42.839\\
53404.1679 & 44.936 & 53488.9726 & 1.966 & 53862.0859 & 48.013 \\
53405.1640 & 0.053 & 53488.9804 & 2.169 &  53862.0976 & 48.259 \\
53405.1757 & 0.000 & 53488.9921 & 2.576  & 53886.0039 & 29.925 \\
53406.1718 & 42.776 & 53490.9609 & 17.779   & 53886.0117 & 30.520 \\
\\
53452.0468 & 44.058 &53490.9726 & 17.187  & 53887.9375 & 0.072 \\
53452.0625 & 44.730 & 53490.9882 & 16.359 & 53887.9492 & 0.103 \\
53453.1171 & 22.795 &53778.1835 & 32.932  & 53887.9570 & 0.016 \\
53453.1328 & 21.890 &53836.1601 & 19.109   & 53890.9531 & 5.842\\
53456.0937 & 7.922 & 53837.0898 & 8.994 & 53891.9140 & 51.517 \\
\\
53456.1054 & 7.464 & 53837.1054 & 9.734 & 53892.0195 & 52.208 \\
53456.1132 & 7.077 & 53859.0742 & 32.515 & 53893.8359 & 14.350 \\
53457.0585 & 22.815 & 53859.0859 & 33.257 & 53893.9062 & 18.418 \\
53457.0664 & 23.287 & 53860.0156 & 42.659 & 53894.0234 & 25.747 \\
53465.0546 & 17.497 & 53860.0234 & 42.385 & 53894.0351 & 26.312 \\
\\
53466.0195 & 50.710 & 53860.9179 & 0.289 & 53894.0429 & 26.808 \\
53467.0429 & 2.357 & 53860.9257 & 0.187  & 53894.7890 & 51.807 \\
53487.0039 & 39.207 & 53861.0312 & 0.083 & 53894.8476 & 50.718 \\
53487.0156 & 39.737 & 53861.0390 & 0.036 & 53894.8593 & 50.569 \\
\\
53894.8710 & 50.161 & 53895.0234 & 45.232 & 53929.8789 & 51.087\\
53894.9414 & 48.181 & 53927.8437 & 20.216 & &\\
53895.0117 & 45.699 & 53929.8632 & 51.420& &\\
\end{tabular}
\end{table*}
\end{center}

\renewcommand{\arraystretch}{0.8}
\begin{center}
\begin{table*}

\caption[Radial velocity of HD\,136905]{Radial velocities of
HD\,136905 relative to the image recorded at JD2453452.105.}
\label{rv:HD136905}

\begin{tabular}{r r r r r r}

\hline
\textbf{HJD } & \textbf{$V_\mathrm{rad}$} & \textbf{HJD } & \textbf{$V_\mathrm{rad}$} & \textbf{HJD } & \textbf{$V_\mathrm{rad}$} \\
(2400000+) & (km\,\ps) & (2400000+) & (km\,\ps) & (2400000+) & (km\,\ps) \\
\hline

53452.105 & 0.000 & 53859.125 & $-$32.892 & 53886.043 & $-$6.659 \\
53465.082 & 27.549 & 53860.043 & $-$45.228 & 53886.059 & $-$6.344 \\
53466.039 & 27.124 & 53860.059 & $-$45.379 & 53887.973 & 26.000 \\
53487.047 & 25.289 & 53860.078 & $-$45.519 & 53887.984 & 26.078 \\
53487.059 & 25.510 & 53860.094 & $-$45.646 & 53888.000 &  26.227 \\
\\
53487.074 & 25.598 & 53861.094 & $-$48.222 & 53888.012 & 26.314 \\
53487.090 &  25.709 & 53861.105 & $-$48.221 & 53892.035 & $-$23.436 \\
53487.109 & 25.846 & 53861.117 & $-$48.171 & 53892.051 & $-$23.800 \\
53488.070 &  28.160 & 53861.129 & $-$48.129 & 53893.922 & $-$48.124 \\
53489.070 &  20.775 & 53862.109 & $-$39.495 & 53893.934 & $-$48.146 \\
\\
53489.090 &  20.527 & 53862.125 & $-$39.336 & 53893.949 & $-$48.200 \\
53491.008 & $-$19.096 & 53862.137 & $-$39.161 & 53894.961 & $-$45.697 \\
53491.027 & $-$19.506 & 53862.152 & $-$38.919 & 53894.973 & $-$45.582 \\
53859.109 & $-$32.577 & 53886.027 & $-$6.954 & 53927.883 & $-$48.384 \\

\end{tabular}
\end{table*}
\end{center}

\newpage

\renewcommand{\arraystretch}{0.8}
\begin{center}
\begin{table*}

\caption[Radial velocity of HD\,194215]{Radial velocities of
HD\,194215 relative to the image recorded at JD2453666.8515.}
\label{rv:HD194215}

\begin{tabular}{r r r r r r}

\hline
\textbf{HJD } & \textbf{$V_\mathrm{rad}$} & \textbf{HJD } & \textbf{$V_\mathrm{rad}$} & \textbf{HJD } & \textbf{$V_\mathrm{rad}$} \\
(2400000+) & (km\,\ps) & (2400000+) & (km\,\ps) & (2400000+) & (km\,\ps) \\
\hline

53251.9062 & 1.647 & 53666.8398 & 0.004  & 53888.0976 & $-$23.126 \\
53300.8750 & $-$0.920 & 53666.8515 & 0.000  & 53888.1093 & $-$23.127 \\
53303.8984 & $-$1.299 & 53666.8593 & $-$0.004 & 53888.1210 & $-$23.113 \\
53323.8671 & $-$4.016 & 53666.9257 & 0.010 & 53888.1289 & $-$23.121 \\
53457.2148 & $-$25.832 & 53666.9414 & 0.025 & 53888.1367 & $-$23.119 \\
\\
53457.2265 & $-$25.836 & 53667.9257 & $-$0.077 & 53888.1484 & $-$23.119 \\
53457.2382 & $-$25.837 & 53830.2382 & $-$25.566 & 53891.1914 & $-$22.494 \\
53457.2460 & $-$25.836 & 53837.1914 & $-$26.104 & 53891.2031 & $-$22.487 \\
53466.1562 & $-$26.392 & 53837.2070 & $-$26.112 & 53891.2187 & $-$22.487 \\
53489.2148 & $-$26.179 & 53837.2187 & $-$26.111 & 53891.2343 & $-$22.475 \\
\\
53489.2304 & $-$26.185 & 53837.2343 & $-$26.102 & 53891.9843 & $-$22.310 \\
53489.2421 & $-$26.184 & 53859.2109 & $-$26.405 & 53891.9960 & $-$22.303 \\
53491.2109 & $-$26.029 & 53859.2187 & $-$26.402 & 53892.0039 & $-$22.300 \\
53491.2226 & $-$26.028 & 53859.2265 & $-$26.400 & 53894.1093 & $-$21.846 \\
53491.2304 & $-$26.023 & 53859.2343 & $-$26.407 & 53894.1171 & $-$21.839 \\
\\
53607.8906 & 0.083 & 53859.2421 & $-$26.394 &  53894.1250 & $-$21.840\\
53607.9023 & 0.087 & 53859.2539 & $-$26.395 &  53894.1328 & $-$21.840\\
53608.0117 & 0.099 & 53860.2109 & $-$26.344 & 53895.0390 & $-$21.640 \\
53608.0195 & 0.102 & 53860.2226 & $-$26.351 & 53895.0507 & $-$21.642 \\
53638.9843 & 1.517 & 53860.2304 & $-$26.346 & 53928.0000 & $-$12.513 \\
\\
53638.9921 & 1.507 & 53860.2382 & $-$26.343 & 53928.0117 & $-$12.426 \\
53639.0000 & 1.483 & 53861.0546 & $-$26.298 & 53928.0195 & $-$12.442 \\
53639.9101 & 1.544 & 53861.0625 & $-$26.323 & 53928.0312 & $-$12.428 \\
53639.9218 & 1.548 & 53861.0742 & $-$26.321 & 53929.0820 & $-$12.134 \\
53641.8085 & 1.518 & 53861.0820 & $-$26.327 & 53929.0937 & $-$12.133 \\
\\
53641.8203 & 1.520 & 53862.0585 & $-$26.256 & 53929.1054 & $-$12.139 \\
53641.8281 & 1.511 & 53862.0664 & $-$26.287 & 53999.8320 & 1.476 \\
\\
53999.8476 & 1.479 & 54000.9101 & 1.548 & 54012.8867 & 1.733 \\
53999.8671 & 1.488 & 54000.9257 & 1.557 & 54014.8085 & 1.720 \\
53999.8867 & 1.489 & 54000.9375 & 1.555 & 54014.8203 & 1.735 \\
53999.9023 & 1.485 & 54000.9492 & 1.544 & 54014.8242 & 1.755 \\
53999.9179 & 1.471 & 54012.8632 & 1.767 &  \\
53999.9335 & 1.478 & 54012.8750 & 1.744 &  \\

\end{tabular}
\end{table*}
\end{center}




\end{document}